# Hybrid algorithms to solve linear systems of equations with limited qubit resources

Fang Gao[a], Guojian Wu[a], Mingyu Yang[a], Wei Cui[b], Feng Shuang[a,*]

[a] *School of Electrical Engineering, Guangxi University, Nanning 530004, China*

[b] *School of Automation Science and Engineering, South China University of Technology, Guangzhou 510640, China*

**Abstract**

The solution of linear systems of equations is a very frequent operation and thus important in many fields. The complexity using classical methods increases linearly with the size of equations. The HHL algorithm proposed by Harrow et al. achieves exponential acceleration compared with the best classical algorithm. However, it has a relatively high demand for qubit resources and the solution $|x\rangle$ is in a normalized form. Assuming that the eigenvalues of the coefficient matrix of the linear systems of equations can be represented perfectly by finite binary number strings, three hybrid iterative phase estimation algorithms (HIPEA) are designed based on the iterative phase estimation algorithm in this paper. The complexity is transferred to the measurement operation in an iterative way, and thus the demand of qubit resources is reduced in our hybrid algorithms. Moreover, the solution is stored in a classical register instead of a quantum register, so the exact unnormalized solution can be obtained. The required qubit resources in the three HIPEA algorithms are different. HIPEA-1 only needs one single ancillary qubit. The number of ancillary qubits in HIPEA-2 is equal to the number of nondegenerate eigenvalues of the coefficient matrix of linear systems of equations. HIPEA-3 is designed with a flexible number of ancillary qubits. The HIPEA algorithms proposed in this paper broadens the application range of quantum computation in solving linear systems of equations by avoiding the problem that quantum programs may not be used to solve linear systems of equations due to the lack of qubit resources.

* This work was supported by the National Natural Science Foundation of China under Grants 61773359, 61720106009 and 61873317.

* Corresponding author at: School of Electrical Engineering, Guangxi University, Nanning 530004, China

*E-mail addresses:* fgao@gxu.edu.cn (F. Gao), gjwu@st.gxu.edu.cn (G. Wu), ymyliangzi@163.com (M. Yang), aucuiwei@scut.edu.cn (W. Cui), fshuang@gxu.edu.cn (F. Shuang).

## 1. Introduction

The solution of linear systems of equations $Ax=b$ is one basic problem in linear algebra, and the explosive growth of the problem scale drives us to develop faster and more efficient algorithms[1]. The quantum HHL algorithm proposed by Harrow, Hassidim and Lloyd[2] can achieve exponential speedup compared to the traditional conjugate gradient algorithm[3]. However, the number of qubits required in the HHL algorithm increases with the number of the linear systems of equations and the magnitude of eigenvalues of the coefficient matrix $A$ [3–7]. In the Noisy Intermediate-Scale Quantum Era (NISQ Era), qubit resources are limited[8] (i.e. the number of qubits is between 50 and 100, and the quantum gates are not perfect). Therefore, the development of algorithms to solve linear systems of equations with lower requirements for qubit resources is of practical significance and application value.

In the exploration of reducing the quantum circuit depth, Lee et al. proposed a hybrid HHL algorithm[9] mainly composed of phase estimation[10–15], classical computing and reduced HHL algorithm, and its core is to process the information extracted from the phase estimation in a classical computer to design the quantum circuit of reduced HHL algorithm. When the coefficient matrix $A$ meets certain conditions, the depth of quantum circuit can be reduced, which is favored due to imperfect quantum gates in the NISQ Era. In this work, a hybrid framework combining quantum and classical approaches is proposed to further reduce the required number of qubits when solving linear systems of equations.

Phase estimation is an important module in the HHL algorithm. Dobšíček et al. proposed an iterative phase estimation algorithm (IPEA), which improves the accuracy of phase estimation by increasing the number of iterations instead of that of ancillary qubits[16–19]. In this work, three hybrid iterative phase estimation algorithms (HIPEA) based on IPEA are presented to solve linear systems of equations, and the number of ancillary qubits can be chosen flexibly according to problem size and available qubit resources.

The rest of the paper is organized as follows: Sec. 2 introduces the theoretical basis of HIPEA; Then three HIPEA algorithms are described detailly in Sec. 3 followed by the corresponding examples and error analysis in Sec. 4; The final conclusions are given in Sec. 5.

## 2. Theoretical Basis

For linear systems of equations $Ax=b$, it is assumed in this paper that the eigenvalues $\phi_j$ of the matrix $A$ can all be perfectly represented by finite binary number strings (i.e. $\phi_j = 0.\phi_{j1}\phi_{j2}...\phi_{j(m-1)}\phi_{jm}$ ), and $b$ is a normalized vector that can be prepared as $|b\rangle$ with QRAM[20,21] and expressed as linear combinations of $A$'s eigenvectors. In the following, the general idea of IPEA and HIPEA will be introduced.

### 2.1. IPEA

In the following description of IPEA, it is assumed that the matrix $A$ only has one eigenvalue $\phi_1$. The IPEA evaluates the eigenvalue digit by digit at different iteration steps. The quantum circuit of the general *l*-th iteration shown in Fig. 1 is employed to estimate the *k*-th bit (from high to low unless otherwise specified) of $\phi_1$ with $k = m + 1 - l$.

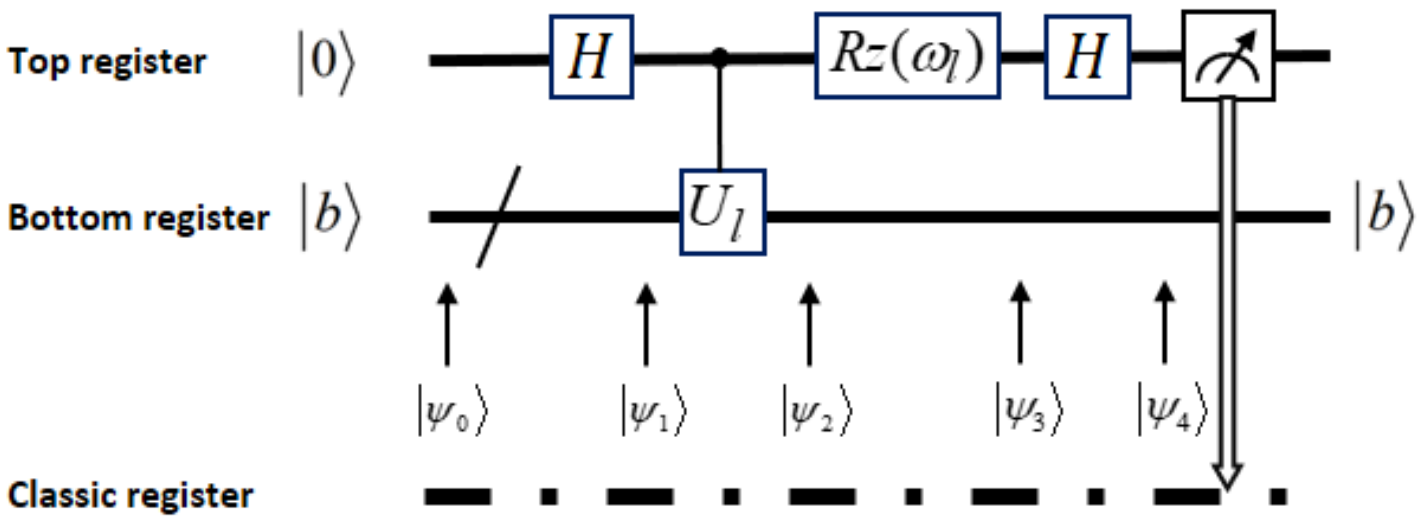

Fig. 1. The *l*-th iteration of the IPEA. The same kind of quantum circuits is adopted in HIPEA.

The evolution of the quantum state in the *l*-th iteration is described in the following. The initial quantum state is $|\psi_0\rangle = |0\rangle\,|b\rangle$. After the first Hadamard gate on the ancillary qubit of the top register, the state becomes $|\psi_1\rangle = 2^{-1/2}(|0\rangle + |1\rangle)|b\rangle$. Then the controlled-U gate with $U_l = U^{2^{m-l}} = e^{2\pi i A\times 2^{m-l}} = e^{2\pi i A\times 2^{k-1}}$ is used to perform the Hamiltonian simulation[22–27] and the resulted state is $|\psi_2\rangle = 2^{-1/2}(|0\rangle + e^{2\pi i(2^{k-1}\phi_1)}|1\rangle)|b\rangle$. The Rz gate, with the angle $\omega_l = \omega_{m+1-k} = 0.0\phi_{1(k+1)}\phi_{1(k+2)}\ldots\phi_{1(m-1)}\phi_{1m}$ determined by the previous iterations, is imposed on the ancillary qubit leading to $|\psi_3\rangle = 2^{-1/2}\times e^{-i\omega_l/2}(|0\rangle + e^{2\pi i(0.\phi_{1k})}|1\rangle)|b\rangle$. After the second Hadamard gate on the ancillary qubit, the state evolves to $|\psi_4\rangle = 2^{-1/2}\times e^{-i\omega_l/2}\left[\left(1+e^{2\pi i(0.\phi_{1k})}\right)|0\rangle + \left(1-e^{2\pi i(0.\phi_{1k})}\right)|1\rangle\right]|b\rangle$. Therefore, the probability of measuring the top register qubit as $|0\rangle$ is $P_0 = \cos^2[\pi(0.\phi_{1k})]$ and that of measuring as $|1\rangle$ is $P_1 = \sin^2[\pi(0.\phi_{1k})]$, which means that the output state of the top register qubit is definitely $|0\rangle$ when $\phi_{1k} = 0$ and $|1\rangle$ when $\phi_{1k} = 1$. So the accurate value of $\phi_1 = 0.\phi_{11}\phi_{12}\ldots\phi_{1(m-1)}\phi_{1m}$ can be evaluated by $m$ iteration steps with IPEA.

### *2.2. HIPEA*

Given linear systems of equations $Ax = b$, the matrix $A$ can be decomposed in its eigenbasis as $A = \Sigma_j \phi_j u_j u_j^{-1}$ with $\phi_j$ as the eigenvalue and $u_j$ as the corresponding normalized eigenvector (i.e. $u_j$ can also be written as $|u_j\rangle$). The vector $b$ can be written as linear combinations of $u_j$ as $b = \Sigma_j \beta_j u_j$ with $\beta_j$ as the projection coefficients. Then the solution of $Ax = b$ is $x = \Sigma_j \beta_j u_j/\phi_j$. IPEA outputs a certain eigenvalue $\phi_j$ of the input matrix $A$ one bit by one bit, and other information in iterations has to be used to solve $Ax = b$. In HIPEA, $|\beta_j|$ and $|u_{jp}|$ ($u_j = [u_{j1}, u_{j2}, \cdots, u_{jp}, \cdots]^T$) can be obtained by quantum measurements in the iterative process with quantum circuits inspired by IPEA. According to Lemma 1, after $m$ iterations, the probability of obtaining the eigenvalue $\phi_j = 0.\phi_{j1}\phi_{j2}\cdots\phi_{jm}$ in the classic register is $\beta_j^2$. Through the post-selection operation[28], when $\phi_j$ is obtained in the $m$ iterations, $u_{jp}^2$ can be obtained as the probability of the corresponding product state of the bottom register (i.e. $u_{j1}^2$ is the probability of $|00\cdots 0\rangle$, $u_{j2}^2$ is the probability of $|00\cdots 1\rangle$). Finally, the overall sign information of $\beta_j u_{jp}$ can be judged with the help of $b = \Sigma_j \beta_j u_j$, and thus the solution $x = \Sigma_j \beta_j u_j/\phi_j$ can be obtained.

In the design of our hybrid algorithms, it is assumed that the matrix has $N$ different eigenvalues. Both IPEA and HIPEA require two quantum registers (top and bottom registers) to execute quantum operations and a classic register to store measurement results. It is worth mentioned the number of ancillary qubits in the top register can be different in our developed HIPEA algorithms, which is described in details in Sec. 3.

In the following, two lemmas concerning the information extraction and the accurate solution of $Ax=b$ are introduced together with proofs as the theoretical basis of HIPEA.

**Lemma 1.** If the eigenvalue $\phi_{j'}$ differs from any other eigenvalue $\phi_{j_{no}}$, where $j_{no}$ is an integer ranging from 1 to $N$ except $j'$, the probability of obtaining the eigenvalue $\phi_{j'}=0.\phi_{j'1}\phi_{j'2}...\phi_{j'm}$ after *m* iterations is $\beta_{j'}^2$.

**Proof.** The original IPEA algorithm only considers the case of one single eigenvalue. However, the general case is that the matrix $A$ with dimension $N$ has $N$ different eigenvalues $\phi_j$ with the integer $j$ ranging from 1 to $N$, and their corresponding normalized eigenvectors can be written as $|u_j\rangle$. The normalized vector $|b\rangle$ can be decomposed in the eigenspace of $A$ as

$$|b\rangle=\Sigma_j\beta_j|u_j\rangle \tag{1}$$

In case of multiple eigenvalues, the system state $|\psi_4\rangle$ before measurement in Fig. 1 is:

$$|\psi_4\rangle=\sum_{j=1}^{N}\frac{e^{-\frac{i\omega_l}{2}}}{2}\left[\beta_j(1-e^{2\pi i(0.\phi_{jk})})|0\rangle+\beta_j(1+e^{2\pi i(0.\phi_{jk})})|1\rangle\right]|u_j\rangle \tag{2}$$

It is assumed in the following that the target eigenvalue to be estimated is $\phi_{j'}=0.\phi_{j'1}\phi_{j'2}\cdots\phi_{j'(m-1)}\phi_{j'm}$ and its corresponding eigenvector is $|u_{j'}\rangle$, the projection coefficient of $|b\rangle$ on $|u_{j'}\rangle$ is $\beta_{j'}$. Same as in IPEA, the rotation parameter $\omega_l$ in the *l*-th iteration of HIPEA is designed to be $\omega_l=0.0\phi_{j'(k+1)}...\phi_{j'(m-1)}\phi_{j'm}$ .

In the first iteration (i.e. $l=1$), the probability of measuring as $|0\rangle$ is

$$P_{j'0_iter\ 1}=\sum_{j=1}^{N}\beta_j^2\cos^2[\pi(0.\phi_{jm})] \tag{3}$$

Here in the subscript, $j'$ means that $\phi_{j'}$ is the target eigenvalue, 0 means the ancillary qubit is measured to be $|0\rangle$, and $iter\ 1$ denotes the first iteration. The probability of measuring as $|1\rangle$ is

$$P_{j'0_iter\ 1}=\sum_{j=1}^{N}\beta_j^2\sin^2[\pi(0.\phi_{jm})] \tag{4}$$

In the second iteration, the probabilities of measuring as $|0\rangle$ and $|1\rangle$ are, respectively,

$$P_{j'0_iter\ 2}=\sum_{j=1}^{N}\beta_j^2\cos^2[\pi(0.\phi_{j(m-1)}\phi_{jm}-0.0\phi_{j'm})] \tag{5}$$

$$P_{j'1_iter\ 2}=\sum_{j=1}^{N}\beta_j^2\sin^2[\pi(0.\phi_{j(m-1)}\phi_{jm}-0.0\phi_{j'm})] \tag{6}$$

Generally, the probability of measuring as $|0\rangle$ and $|1\rangle$ in the *l*-th iteration are, respectively,

$$P_{j'0_iter\ l}=\sum_{j=1}^{N}\beta_j^2\cos^2[\pi(0.\phi_{j(m-l+1)}\phi_{j(m-l+2)}...\phi_{j(m-1)}\phi_{jm}-0.0\phi_{j'(m-l+2)}...\phi_{j'(m-1)}\phi_{j'm})] \tag{7}$$

$$P_{j'1_iter\ l}=\sum_{j=1}^{N}\beta_j^2\sin^2[\pi(0.\phi_{j(m-l+1)}\phi_{j(m-l+2)}...\phi_{j(m-1)}\phi_{jm}-0.0\phi_{j'(m-l+2)}...\phi_{j'(m-1)}\phi_{j'm})] \tag{8}$$

The projections of $P_{j'0_iter\ l}$ and $P_{j'1_iter\ l}$ on $|u_j\rangle$ are, respectively,

$$P_{j'0_iter\ l_u_j} = \beta_j^2 \cos^2[\pi(0.\phi_{j(m-l+1)}\phi_{j(m-l+2)}...\phi_{j(m-1)}\phi_{jm} - 0.0\phi_{j'(m-l+2)}...\phi_{j'(m-1)}\phi_{j'm})] \quad (9)$$

$$P_{j'1_iter\ l_u_j} = \beta_j^2 \sin^2[\pi(0.\phi_{j(m-l+1)}\phi_{j(m-l+2)}...\phi_{j(m-1)}\phi_{jm} - 0.0\phi_{j'(m-l+2)}...\phi_{j'(m-1)}\phi_{j'm})] \quad (10)$$

Then the total projection probability on $|u_j\rangle$ of obtaining $\phi_{j'}$'s lowest $l$ bits in $l$ iterations in the classic register is multiplication of the corresponding probabilities in Eqs. (9) and (10) in different iterative steps:

$$P^{total}_{j'\phi_{j'}_iter\ l_u_j} = \beta_j^2 \prod_{l=1}^{m} \frac{P_{j'\phi_{j'(m-l+1)}_iter\ l_u_j}}{\beta_j^2} \quad (11)$$

The total probability of obtaining $\phi_{j'(m-l+1)}\phi_{j'(m-l+2)}...\phi_{j'(m-1)}\phi_{j'm}$ in $l$ iterations in the classic register is

$$P^{total}_{j'\phi_{j'}_iter\ l} = \sum_{j=1}^{N} P^{total}_{j'\phi_{j'}_iter\ l_u_j} \quad (12)$$

Without losing generality, it is assumed that the matrix $A$ has eigenvalues $\phi_{j'}$ and $\phi_{j_{no}}$ with the low $m-d$ bits of $\phi_{j'}$ the same as those of $\phi_{j_{no}}$ and $\phi_{j'd} \neq \phi_{j_{no}d}$. Then in the $(m-d)$-th iteration,

$$P^{total}_{j'\phi_{j'}_iter\ (m-d)_u_{j'}} = \beta_{j'}^2 \quad (13)$$

$$P^{total}_{j'\phi_{j'}_iter\ (m-d)_u_{j_{no}}} = \beta_{j_{no}}^2 \quad (14)$$

In the $(m-d+1)$-th iteration, due to $P_{j'\phi_{j'd}_iter\ (m-d+1)_u_{j'}} = \beta_{j'}^2$ and $P_{j'\phi_{j'd}_iter\ (m-d+1)_u_{j_{no}}} = 0$, we have

$$P^{total}_{j'\phi_{j'}_iter\ (m-d+1)_u_{j'}} = P^{total}_{j'\phi_{j'}_iter\ (m-d)_u_{j'}} \times P_{j'\phi_{j'd}_iter\ (m-d+1)_u_{j'}} / \beta_{j'}^2 = \beta_{j'}^2 \quad (15)$$

$$P^{total}_{j'\phi_{j'}_iter\ (m-d+1)_u_{j_{no}}} = P^{total}_{j'\phi_{j'}_iter\ (m-d)_u_{j_{no}}} \times P_{j'\phi_{j'd}_iter\ (m-d+1)_u_{j_{no}}} / \beta_{j_{no}}^2 = 0 \quad (16)$$

The above relations indicate that when $\phi_{j_{no}}$ is different from $\phi_{j'}$, there must exist an iteration (i.e. $m-d+1$) making the probability

$$P_{j'\phi_{j'd}_iter\ (m-d+1)_u_{j_{no}}} = 0 \quad (17)$$

It is obvious that the parameter $d$ is determined by bit location where $\phi_{j'}$ and $\phi_{j_{no}}$ differ at. When the eigenvalue $\phi_{j'}$ differs from any other eigenvalue $\phi_{j_{no}}$ for the low $m-d+1$ bits, where $j_{no}$ is an integer ranging from 1 to $N$ except $j'$, we can draw the following conclusion and complete the proof according to Eq. (12)

$$\begin{aligned} P^{total}_{j'\phi_{j'}_iter\ m} &= P^{total}_{j'\phi_{j'}_iter\ m-d+1} \\ &= \sum_{j=1}^{j'-1} P^{total}_{j'\phi_{j'}_iter\ m-d+1_u_j} + P^{total}_{j'\phi_{j'}_iter\ m-d+1_u_{j'}} + \sum_{j=j'+1}^{N} P^{total}_{j'\phi_{j'}_iter\ m-d+1_u_j} \\ &= 0 + \beta_{j'}^2 + 0 = \beta_{j'}^2 \end{aligned} \quad (18)$$

**Lemma 2.** Different from the HHL algorithm, HIPEA can obtain the accurate solution $x$ instead of its normalized vector form $|x\rangle$.

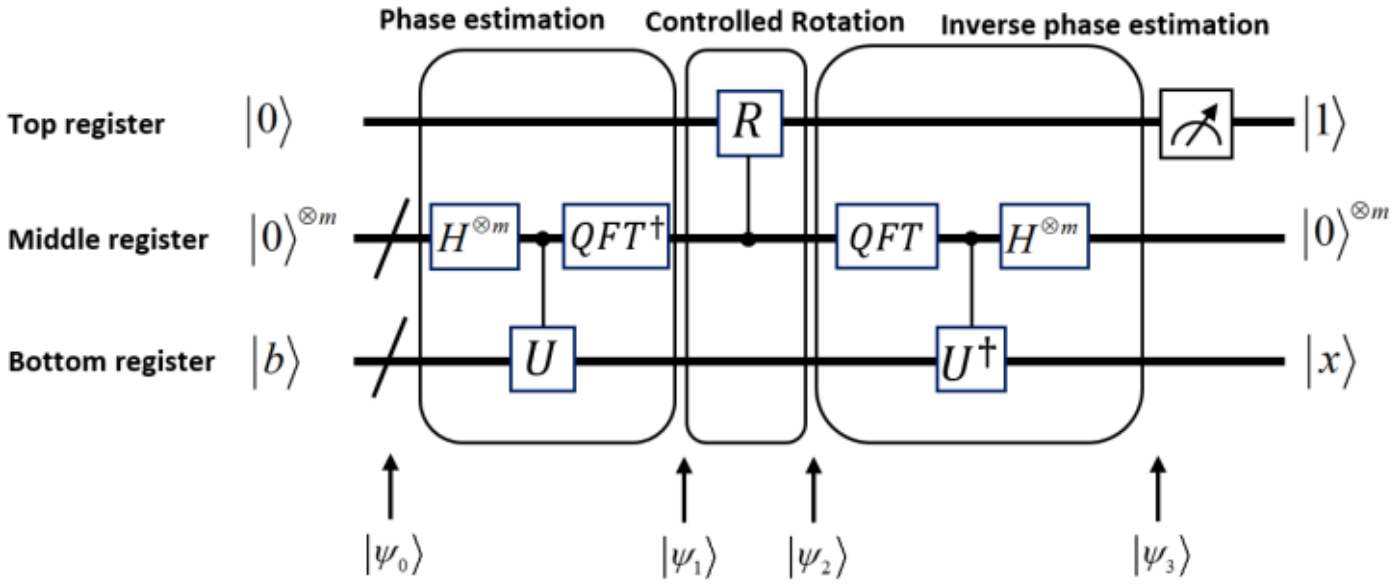

Fig. 2 Quantum circuit of the HHL algorithm.

**Proof.** Fig. 2 shows the quantum circuit of the original HHL algorithm to extract $|x\rangle$, and it consists of the following steps: (1) The initial state of the system is prepared to be $|\psi_0\rangle = |0\rangle\,|0\rangle^{\otimes m}|b\rangle$. (2) After the phase estimation module, the state becomes $|\psi_1\rangle = \Sigma_{j=1}^{N}\beta_j|\phi_j\rangle\,|u_j\rangle$. (3)The controlled rotation operation makes the state evolve to $|\psi_2\rangle = \Sigma_{j=1}^{N}\beta_j(\sqrt{1-C^2/\phi_j^2}\,|0\rangle + C/\phi_j|1\rangle)|\phi_j\rangle\,|u_j\rangle$, where $C$ is a normalizing constant chosen to ensure rotations are less than $2\pi$ [29]. (4) The inverse phase estimation makes the qubits in the middle register unentangled with those in the other two registers, and the output state is $|\psi_3\rangle = \Sigma_{j=1}^{N}\beta_j(\sqrt{1-C^2/\phi_j^2}\,|0\rangle + C/\phi_j|1\rangle)|0\rangle^{\otimes m}|u_j\rangle$. (5) When the ancillary qubit in the top register is measured as $|1\rangle$, the qubits of the bottom register will collapse to the state $|x\rangle = C\Sigma_{j=1}^{N}(\beta_j/\phi_j)|u_j\rangle$ [5,29,30].

As seen in the above procedures, a normalization constant $C$ is introduced in order to normalize $|x\rangle$, which means that an additional step is required to extract $C$ to obtain the accurate solution $x = |x\rangle/C = \Sigma_{j=1}^{N}(\beta_j/\phi_j)|u_j\rangle$.

Different from HHL, the extracted information ($\phi_j$, $|\beta_j|$ and $|u_{jp}|$) for solving the linear systems of equations is stored in the classic register without normalization constraint. Theoretically, the exact values of $\phi_j$ and $\beta_j u_j$ can be extracted within a certain error range, and thus the accurate solution $x = \Sigma_j \beta_j u_j/\phi_j$ can be directly obtained. Therefore, the proof is completed.

## 3. Three HIPEA Algorithms

HIPEA is a hybrid framework combining quantum and classical approaches, and the quantum part is performed on quantum circuits, whose design are constrained by available qubit resources (i.e. the number of qubits). In this section, three different HIPEA algorithms will be given in details, and they can be chosen flexibly according to available qubit resources.

Without losing generality, the $N$ different eigenvalues $\phi_j = 0.\phi_{j1}\phi_{j2}\cdots\phi_{jm}$ of the coefficient matrix $A$ in $Ax = b$ can be described by a binary tree shown in Fig. 3(a), where $\phi_{j(m+1-l)}$ represents the estimated $(m+1-l)$-th bit of $\phi_j$ in the *l*-th iteration, $\phi_{\{1,2,\cdots,N\}}$ represents the $N$ eigenvalues (i.e. $\phi_1, \phi_2, \cdots, \phi_N$), and $\phi_{\{1,2,\cdots,N\}k}$ represents their *k*-th bit.

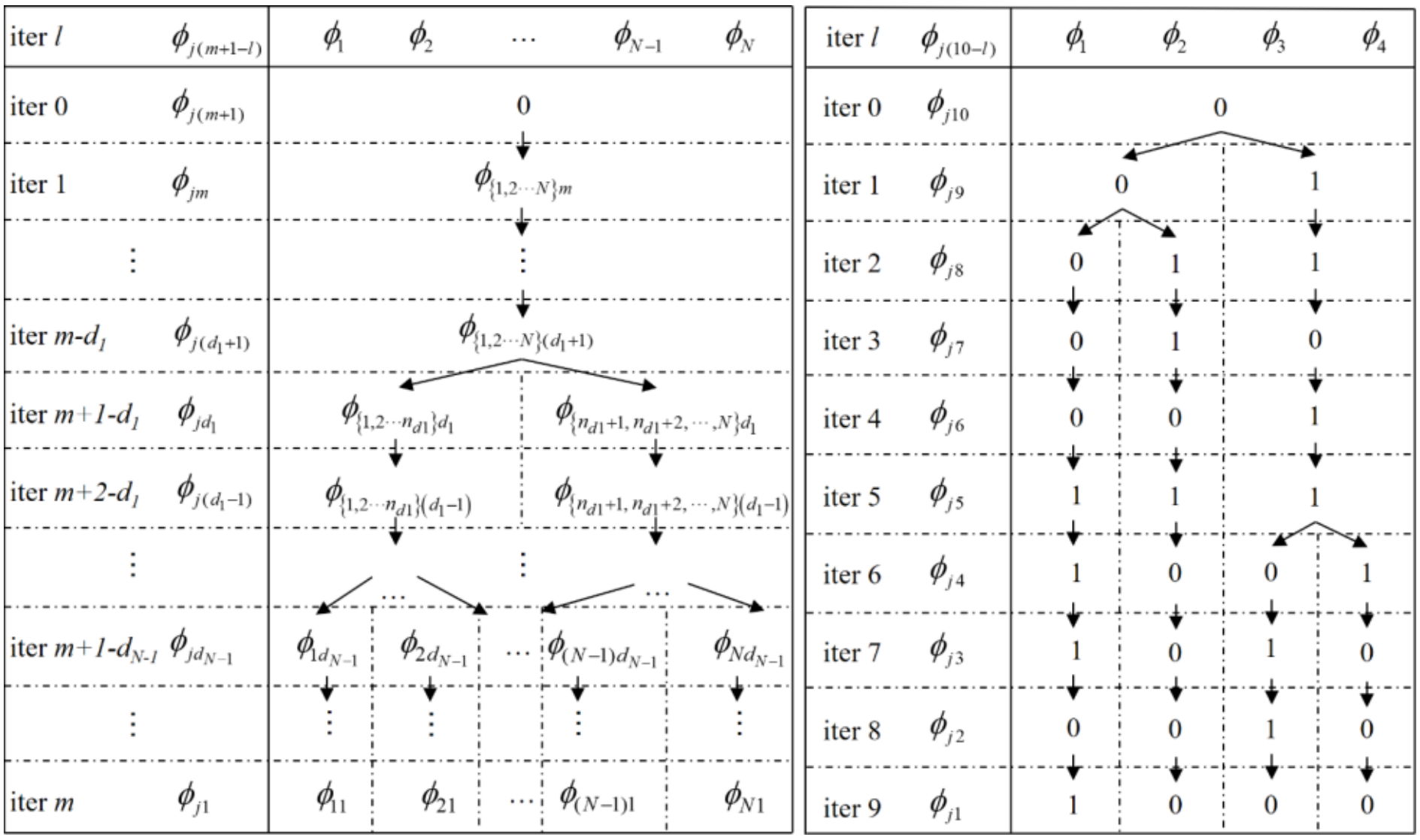

(a) A binary tree representation of $N$ different eigenvalues $\phi_j$ (b) An example binary tree of four eigenvalues

Fig.3 Binary tree representations of eigenvalues

If the low $m-d_1$ bits for all eigenvalues $\phi_j$ are the same (i.e. $\phi_{j(d_1+1)}\phi_{j(d_1+2)}\cdots\phi_{jm}$ are the same for all $j$) and the first divergence appears at the $d_1$-th bit of $\phi_j$, then in the $(m+1-d_1)$-th iteration, the eigenvalues can be divided into two groups with the $n_{d1}$-th eigenvalue $\phi_j$ as the boundary

$$\phi_{\{1,2\cdots n_{d1}\}d_1} \neq \phi_{\{n_{d1}+1,\ n_{d1}+2,\ \cdots\ ,N\}d_1} \tag{19}$$

Similarly, after the $(t-1)$-th divergence appears in the $(m+1-d_{t-1})$-th iteration, the eigenvalues can be divided into $t$ groups according to their low $m+1-d_{t-1}$ bits. Until the $(N-1)$-th divergence appeared in the $(m+1-d_{N-1})$-th iteration, all the $N$ different eigenvalues are differentiated.

One specific example is given in Fig. 3(b), where there are four different eigenvalues:

$$\begin{aligned} &\phi_1 = 0.101110000,\quad \phi_2 = 0.000010110 \\ &\phi_3 = 0.011011011,\quad \phi_4 = 0.000111011 \end{aligned} \tag{20}$$

The first divergence appears in the first iteration resulting to two groups $\phi_{\{1,2\}}$ and $\phi_{\{3,4\}}$; The second divergence appears in the second iteration, leading to $\phi_1$, $\phi_2$ and $\phi_{\{3,4\}}$; The third divergence in the sixth iteration differentiates all the four eigenvalues.

The following three HIPEA algorithms will be described based on the concept of "differentiation at different divergences", and the algorithms include four steps: (1)Extract the eigenvalues $\phi_j$ and the absolute values of the corresponding projection coefficients $|\beta_j|$; (2)Extract the absolute value $|u_{jp}|$ of each element of the normalized eigenvector; (3)Extract the values of $\beta_j u_j$; (4)Obtain the solution of linear systems of equations with the above extracted information.

The difference of the following three HIPEA algorithms mainly lies in the scheme of extracting $\phi_j$ and $|\beta_j|$ in the first step.

### *3.1. HIPEA-1 with single ancillary qubit*

Among the three HIPEA algorithms, HIPEA-1 requires the least number of ancillary qubits. The eigenvalue extraction process of HIPEA-1 is almost the same as that of IPEA except the design of rotation parameters. As described in Sec. 2.1, IPEA only considers the case of one single eigenvalue, and it always collapse to $|0\rangle$ or $|1\rangle$ with a probability of $1$ when the ancillary qubit in the top register is measured in each iteration. While HIPEA-1 considers multiple eigenvalues, and the state of the ancillary qubit collapses to $|0\rangle$ or $|1\rangle$ with the probability no longer necessarily being $0$ or $1$ after measurement according to Eqs. (7) and (8).

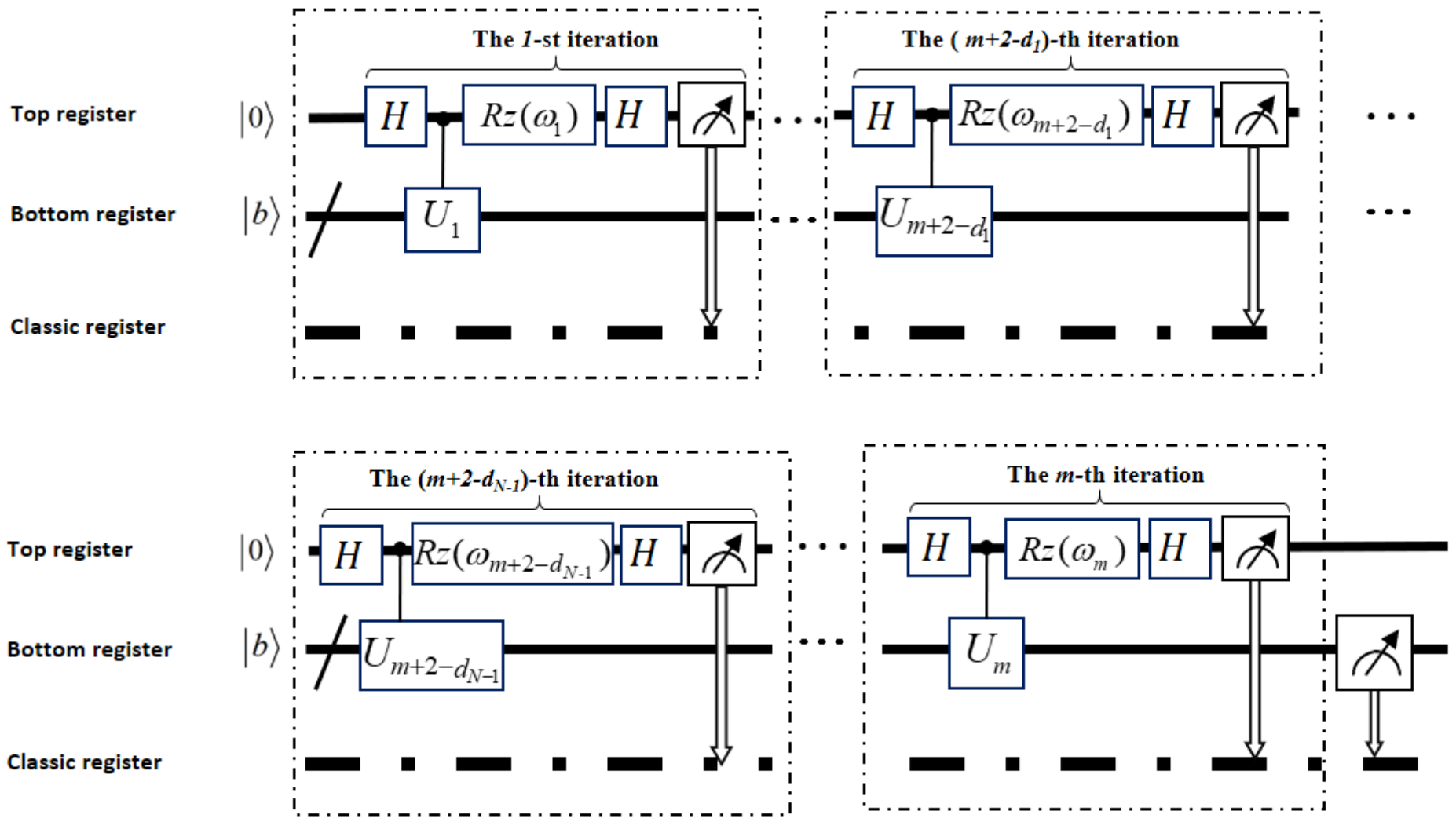

Fig.4 Quantum circuits of HIPEA-1 in different iterations

Taking the case in Fig. 3(a) as an example, Fig. 4 shows quantum circuits of HIPEA-1 in different iterations, where the top register, the unitary matrix $U_l$ and rotation parameters $\omega_l$ are the same as in IPEA described in Sec. 2.1. Unless otherwise stated, the states of qubits in the top and bottom registers are, respectively, initialized to $|0\rangle$ and $|b\rangle$ after each iteration.

The four steps of HIPEA-1 are as follows:

(1) Extract the eigenvalues $\phi_j$ and the absolute values of the corresponding projection coefficients $|\beta_j|$.

In the first iteration shown in Fig. 4, the rotation parameter of the Rz gate is $\omega_1 = -2\pi(0.0)$. The probability of the ancillary qubit being measured as $|0\rangle$ in the top register is $P_{j'0_iter\,1}$, and the probability of being measured as $|1\rangle$ is $P_{j'1_iter\,1}$. The rotation parameter $\omega_2$ in the second iteration is based on the measurement results of the first iteration. When no divergence appears in the first iteration, we have $P_{j'\phi_{jm}_iter\,1} = 1$ according to Eqs. (7) and (8), and the rotation parameter in the second iteration is $\omega_2 = -2\pi(0.0\phi_{jm})$. The first divergence of different eigenvalues appears in the $d_1$-th bit, which means that after $l$ iterations, the measurement results satisfy

$$\begin{aligned} & P^{total}_{j'\phi_{j'}_iter\,l} = 1 \\ & s.t.\ \ l < m+1-d_1 \end{aligned} \tag{21}$$

In the $(m+1-d_1)$-th iteration, $\phi_{\{1,2\cdots n_{d1}\}d_1} \neq \phi_{\{n_{d1}+1, n_{d1}+2,\cdots,N\}d_1}$, according to Eqs. (7) and (8), the measurement probability of being $|0\rangle$ or $|1\rangle$ is no longer $0$ or $1$:

$$0 < P^{total}_{\{1,\cdots,n_{d1}\}\phi_{\{1,\cdots,n_{d1}\}}_iter\, m+1-d_1} \times P^{total}_{\{n_{d1}+1,\cdots,N\}\phi_{\{n_{d1}+1,\cdots,N\}}_iter\, m+1-d_1} < 1 \tag{22}$$

After the first divergence appears in the $(m+1-d_1)$-th iteration, two experiments are performed in the $(m+2-d_1)$-th iteration: In experiment 1, the rotation parameters in the $(m+2-d_1)$-th iteration is designed as $\omega_{m+2-d_1} = -2\pi(0.0\phi_{\{1,\cdots,n_{d1}\}d_1}\phi_{\{1,\cdots,n_{d1}\}(d_1+1)}\cdots\phi_{\{1,\cdots n_{d1}\}m})$. The probability of obtaining $\phi_{\{1,\cdots,n_{d1}\}d_1}\phi_{\{1,\cdots,n_{d1}\}(d_1+1)}\cdots\phi_{\{1,\cdots,n_{d1}\}m}$ as the measurement results after this iteration is $P^{total}_{\{1,\cdots,n_{d1}\}\phi_{\{1,\cdots,n_{d1}\}}_iter\, l}$; In experiment 2, the rotation parameters in the $(m+2-d_1)$-th iteration is designed as $\omega_{m+2-d_1} = -2\pi(0.0\phi_{\{n_{d1}+1,\cdots,N\}d_1}\phi_{\{n_{d1}+1,\cdots,N\}(d_1+1)}\cdots\phi_{\{n_{d1}+1,\cdots,N\}m})$. The probability of obtaining $\phi_{\{n_{d1}+1,\cdots,N\}d_1}\phi_{\{n_{d1}+1,\cdots,N\}(d_1+1)}\cdots\phi_{\{n_{d1}+1,\cdots,N\}m}$ as the measurement results after this iteration is $P^{total}_{\{n_{d1}+1,\cdots,N\}\phi_{\{n_{d1}+1,\cdots,N\}}_iter\, l}$.

Generally, after the $(t-1)$-th divergence appears in the $(m+1-d_{t-1})$-th iteration, $t$ different experiments are performed in the $(m+2-d_{t-1})$-th iteration. The target eigenvalues in the $(m+2-d_{t-1})$-th iteration of the $t$ experiments are different, so the corresponding rotation parameters $\omega_{m+2-d_{t-1}}$ are also different.

After the $(N-1)$-th divergence appears in the $(m+1-d_{N-1})$-th iteration, $N$ different experiments are performed in the $(m+2-d_{N-1})$-th iteration, with the experiment $j$ taking $\phi_j$ as the target eigenvalue. According to Lemma 1, after the $m$-th iteration, the probability of obtaining $\phi_j$ is $\beta_j^2$. Namely, all the eigenvalues $\phi_j$ of $A$ and the absolute values of the corresponding projection coefficients $|\beta_j|$ can be extracted after $m$ iterations.

(2) Extract the absolute value $|u_{jp}|$ of each element of the normalized eigenvector.

Post-selection is a very common processing method in the field of quantum computing. When the measuring results of the ancillary qubit in the top register are $\phi_{j1}\phi_{j2}\cdots\phi_{j(m-1)}\phi_{jm}$ in the $m$ iterations, $u_{jp}^2$ can be measured as the probabilities of qubits in the bottom register being the Kronecker product basis states. The $|u_{jp}|$ can be calculated as the square root of $u_{jp}^2$.

(3) Extract the values of $\beta_j u_j$.

The following equations can be derived from Eq. (1)

$$\begin{aligned}
&(-1)^{n_{11}} \times |\beta_1 u_{11}| + (-1)^{n_{21}} \times |\beta_2 u_{21}| + \cdots + (-1)^{n_{N1}} \times |\beta_N u_{N1}| = b_1 \\
&(-1)^{n_{12}} \times |\beta_1 u_{12}| + (-1)^{n_{22}} \times |\beta_2 u_{22}| + \cdots + (-1)^{n_{N2}} \times |\beta_N u_{N2}| = b_2 \\
&\qquad\vdots \\
&(-1)^{n_{1N}} \times |\beta_1 u_{1N}| + (-1)^{n_{2N}} \times |\beta_2 u_{2N}| + \cdots + (-1)^{n_{NN}} \times |\beta_N u_{NN}| = b_N
\end{aligned} \tag{23}$$

where $n_{jp}$ is an integer of 0 or 1 indicating the positive or negative sign of $\beta_j u_{jp}$. The information of signs can be determined by traversing all possible combinations of $n_{jp}$ and thus $\beta_j u_j$ can be obtained.

(4) Obtain the solution of linear systems of equations with the above extracted information.

With the extracted $\phi_j$ and $\beta_j u_j$, the solution of the linear systems of equations is obtained as $x = \Sigma_j \beta_j u_j / \phi_j$.

To conclude, in order to obtain the solution of $Ax = b$, we have to conduct different experiments with $m$ iterations to differentiate the eigenvalues of matrix $A$. HIPEA-1 requires the least number of ancillary qubits, that is, only one ancillary qubit is needed in the top register in each iteration. When the $N$ eigenvalues are all non-degenerate, at most $N$ different experiments have to be performed to differentiate these eigenvalues.

### 3.2. HIPEA-2 with $N$ ancillary qubits

HIPEA-2 achieves the same effect with $N$ ancillary qubits instead of $N$ experiments to differentiate $N$ different eigenvalues. The top quantum register in HIPEA-2 contains $N$ ancillary qubits. The quantum circuits of HIPEA-2 in different iterations are shown in Fig. 5.

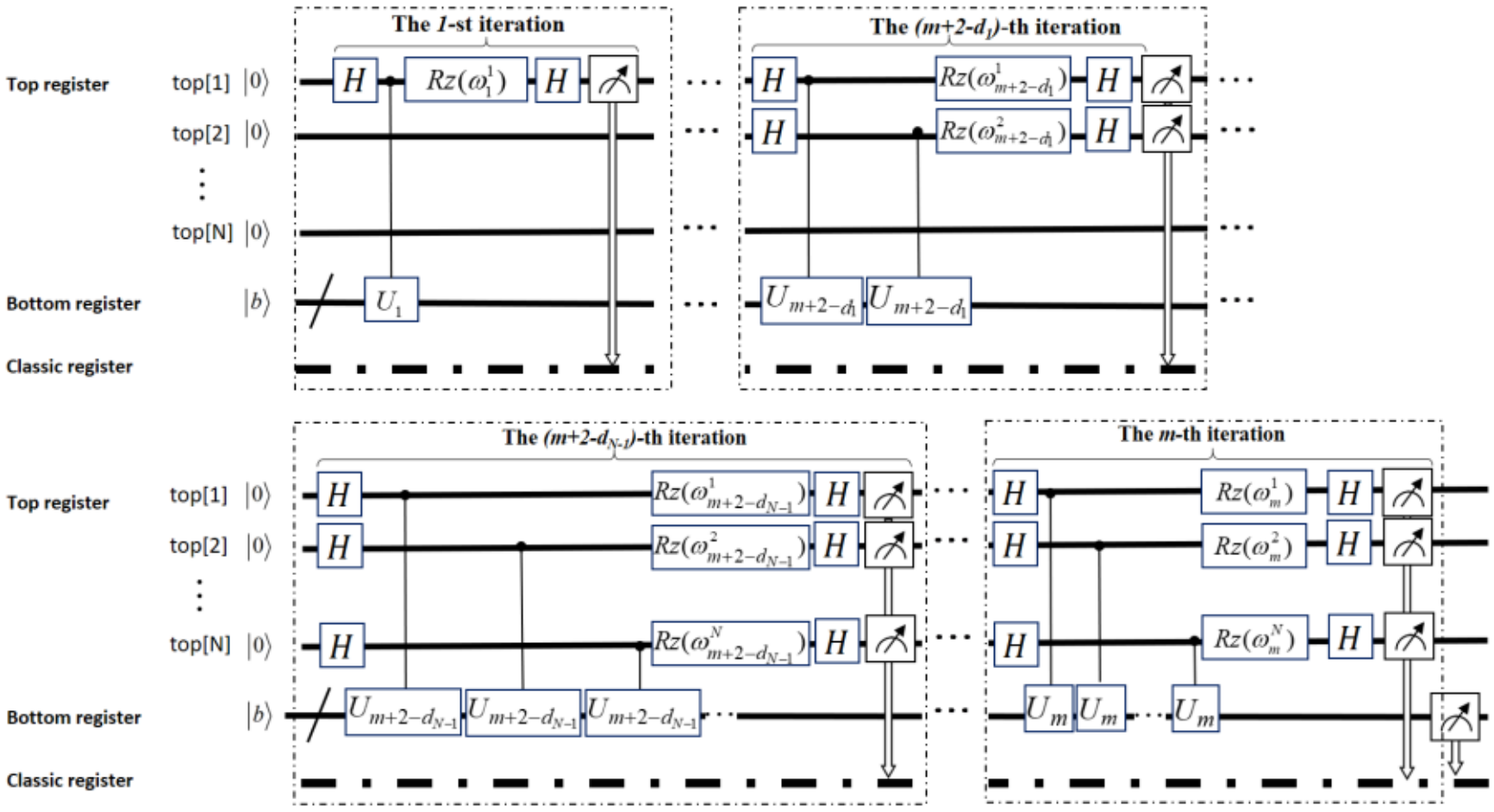

Fig.5 Quantum circuits of HIPEA-2 in different iterations

The unitary matrices $U_l$ and rotation parameters $\omega_l^q$ are designed in the same way as in Sec. 2.1, where the superscript $q$ means that the Rz gate with the rotation parameter $\omega_l^q$ acts on the qubit $top[q]$ in the top register.

The four steps of HIPEA-2 are as follows:

(1) Extract the eigenvalues $\phi_j$ and the absolute values of the corresponding projection coefficients $|\beta_j|$.

In the first iteration shown in Fig. 5, only the qubit $top[1]$ is used in the top register with the rotation parameter $\omega_1^1 = -2\pi(0.0)$. Similar with HIPEA-1, the measurement results are satisfied with Eq. (21) before the first divergence appears. It is worth noted that before the ($m+2-d_1$)-th iteration ($l < m+2-d_1$), no matter which eigenvalue is taken as the target eigenvalue, the corresponding rotation parameters are always the same. Therefore, only one qubit $top[1]$ in the top register is used before the $(m+2-d_1)$-th iteration.

The first divergence appears in the $(m+1-d_1)$-th iteration, and the measurement results in this iteration obeys Eq. (22). The eigenvalues $\phi_{\{1,\cdots,n_{d1}\}}$ and $\phi_{\{n_{d1}+1,\cdots,N\}}$ can be differentiated in the ($m+2-d_1$)-th iteration with the rotation parameter $\omega_{m+2-d_1}$ designed with $\phi_{\{1,\cdots,n_{d1}\}}$ and $\phi_{\{n_{d1}+1,\cdots,N\}}$ as the target eigenvalues, respectively. In HIPEA-2, the same effect can be achieved by increasing the number of qubits used in the top register instead of increasing the number of experiments in HIPEA-1. Namely, in the $(m+2-d_1)$-th iteration in Fig. 5, the rotation parameter $\omega^1_{m+2-d_1}$ of Rz gate in the qubit $top[1]$ is designed with the target eigenvalue $\phi_{\{1,\cdots,n_{d1}\}}$, and the rotation parameter $\omega^2_{m+2-d_1}$ of Rz gate in the qubit $top[2]$ is designed with the target eigenvalue $\phi_{\{n_{d1}+1,\cdots,N\}}$:

$$\omega^1_{m+2-d_1} = -2\pi\left(0.0\phi_{\{1,\cdots n_{d1}\}d_1}\phi_{\{1,\cdots n_{d1}\}(d_1+1)}\cdots\phi_{\{1,\cdots n_{d1}\}m}\right) \tag{24}$$

$$\omega^2_{m+2-d_1} = -2\pi\left(0.0\phi_{\{n_{d1}+1,\cdots,N\}d_1}\phi_{\{n_{d1}+1,\cdots,N\}(d_1+1)}\cdots\phi_{\{n_{d1}+1,\cdots,N\}m}\right) \tag{25}$$

Generally, after the $(t-1)$-th divergence appears in the $(m+1-d_{t-1})$-th iteration, the eigenvalues can be divided into $t$ groups according to their low $m+1-d_{t-1}$ bits. Accordingly, starting from the $(m+2-d_{t-1})$-th iteration, the rotation parameters $\omega_{m+2-d_t-1}$ have $t$ different values, and $t$ qubits in total in the top register are put into use:

$$\omega^1_{m+2-d_{t-1}} \neq \omega^2_{m+2-d_{t-1}} \neq \cdots \neq \omega^t_{m+2-d_{t-1}} \tag{26}$$

where $\omega^q_{m+2-d_{t-1}}$ denotes the rotation parameter of the Rz gate on the qubit $top[q]$.

After the $(N-1)$-th divergence appears in the $(m+1-d_{N-1})$-th iteration, all the $N$ eigenvalues can be differentiated. Accordingly, starting from the $(m+2-d_{N-1})$-th iteration, all the $N$ qubits in the top register are put into use, and the rotation parameter of Rz gate on the qubit $top[q]$ are designed with $\phi_q$ as the target eigenvalue. According to Lemma 1, all the eigenvalues $\phi_j$ and the absolute value of the corresponding projection coefficients $|\beta_j|$ can be extracted after $m$ iterations.

(2) The absolute value $|u_{jp}|$ of each element of the normalized eigenvector is extracted similarly as in HIPEA-1.

(3) The values of $\beta_j u_j$ are extracted similarly as in HIPEA-1.

(4) The solution of the linear systems of equations is obtained as $x = \Sigma_j \beta_j u_j / \phi_j$ .

In HIPEA-2, the number of iterations is the same as in HIPEA-1, but only one experiment has to be conducted instead of $N$ experiments. The complexity just moves from "the number of experiments" to "the number of ancillary qubits", and the number of ancillary qubits in the top register is $N$ when all the eigenvalues of matrix $A$ are not degenerate.

### *3.3. HIPEA-3 with flexible multiple ancillary qubits*

HIPEA-3 tries to differentiate different eigenvalues with flexible multiple ancillary qubits using a traversing strategy. The quantum circuits in different iterations are shown in Fig. 6. The number of ancillary qubits in the top register is $n_{top}$, and the *q*-th qubit is marked as $top[q]$. The meaning of the unitary matrices $U_l$ and rotation parameters $\omega_l$ are the same as in Sec. 2.1.

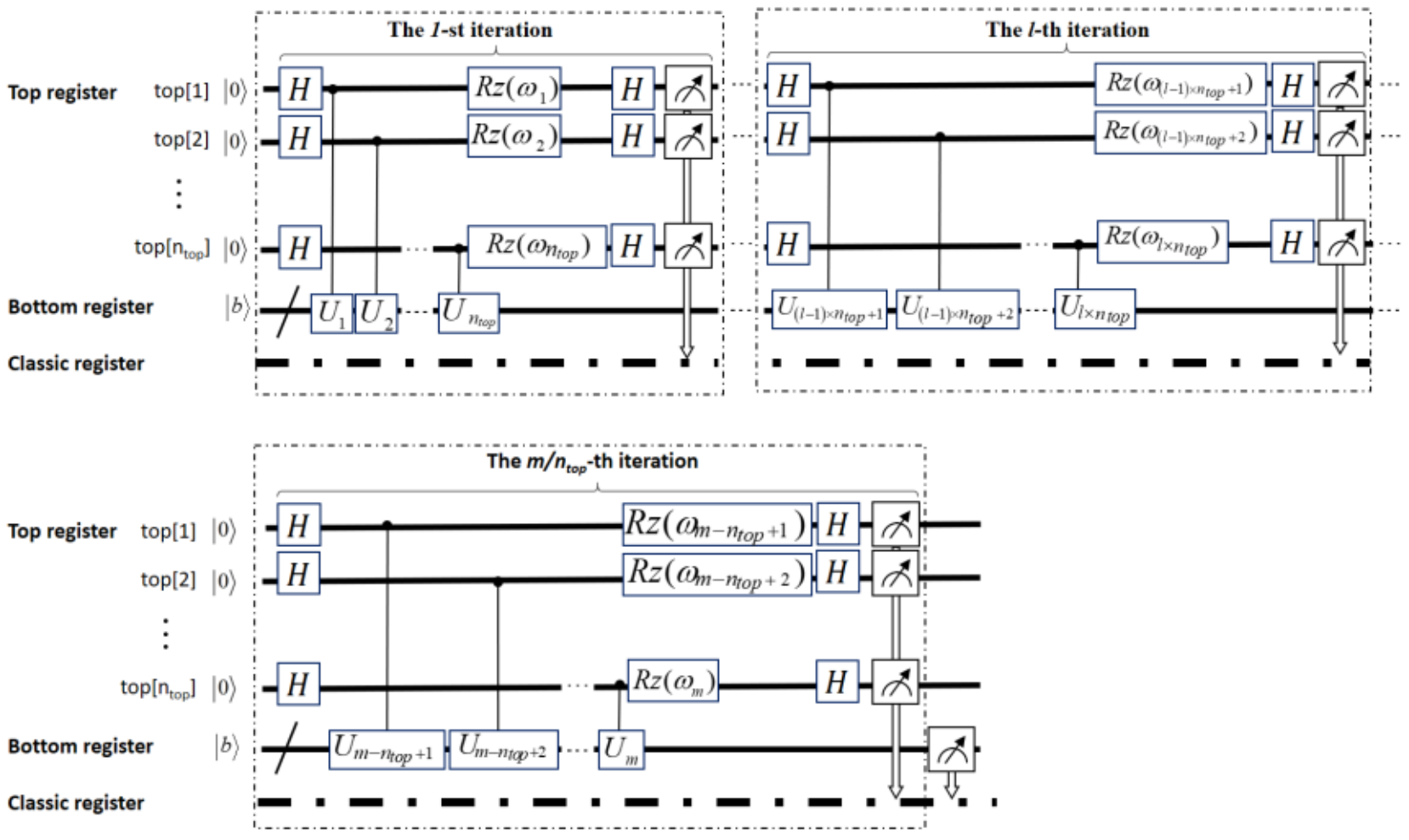

Fig.6 Quantum circuits of HIPEA-3 in different iterations

A set of experiments is performed in each iteration in HIPEA-3 to traverse all possible combinations of rotation parameters $\omega_l$, whereas $\omega_l$ in HIPEA-1 and HIPEA-2 in each iteration are designed based on the measuring results of previous iterations. The operation on the qubit $top[q]$ in the *l*-th iteration in HIPEA-3 can be mapped to the operation on the ancillary qubit in the ($(l-1)\times n_{top}+q$)-th iteration in HIPEA-1, which means that the measuring results of the $l$ iterations in HIPEA-3 can be mapped to those of the $l\times n_{top}$ iterations in HIPEA-1, namely $P^{total}_{j'\phi_{j'}_iter\ l\times n_{top}}$.

Since "traversing" the rotation parameters is equivalent to "traversing" the target eigenvalues, we have $P^{total}_{j'\phi_{j'}_iter\ l\times n_{top}}=0$ in the experiments where the target phase $\phi_{j'}$ is not the eigenvalue of matrix $A$. Non-zero $P^{total}_{j'\phi_{j'}_iter\ l\times n_{top}}$ in HIPEA-3 gives the measuring probability of the low $l\times n_{top}$ bits of the eigenvalue $\phi_{j'}$ of matrix $A$ after *l* iterations. The number of non-zero $P^{total}_{j'\phi_{j'}_iter\ l\times n_{top}}$ after *l* iterations $n^{iter\ l}_{nonzero}$ can be understood as the number of divergences shown in Fig. 3(a) for the low $l\times n_{top}$ bits after *l* iterations.

The rotation parameter of the Rz gate acting on the qubit $top[q]H$ in the *l*-th iteration is denoted as $\omega_{(l-1)\times n_{top}+q}=-2\pi(0.bin^{iter\ l}_q)$, where $bin^{iter\ l}_q$ is a ($q+(l-1)\times n_{top}$)-bit binary number and designed as follows: (1) Similar as the design principle of $\omega_l$ in Sec. 2.1, the highest bit of $bin^{iter\ l}_q$ is always 0; (2) All possible combinations of values of the next $q-1$ bits (from high to low) of $bin^{iter\ l}_q$ are traversed; (3) The low $(l-1)\times n_{top}$ bits of $bin^{iter\ l}_q$ take the same values as the low $(l-1)\times n_{top}$ bits of $\phi_{j'}$ ($P^{total}_{j'\phi_{j'}_iter\ (l-1)\times n_{top}}\neq 0$) extracted after $l-1$ iterations in HIPEA-3.

The four steps of HIPEA-3 are as follows:

(1) Extract the eigenvalues $\phi_j$ and the absolute values of the corresponding projection coefficients $|\beta_j|$:

The goal of the first iteration is to estimate the low $n_{top}$ bits of the eigenvalues $\phi_j$. The rotation parameters of each experiment in iteration 1 are listed in Table 1. Due to the traversal scheme of designing the rotation parameters $\omega_q$, the number of experiments in this iteration is $2^{n_{top}-1}$, and all non-zero measurement results $P^{total}_{j'\phi_{j'}_iter\ n_{top}}$ can be obtained from these experiments.

Table 1 $\omega_{(l-1)\times n_{top}+q}$ for each experiment in the first iteration of HIPEA-3

| | Qubit $top[q]$ | Rotation parameter $\omega_{(l-1)\times n_{top}+q}$ | Experiment number: Experiment 1 | ··· | Experiment $2^{n_{top}-1}$ |
|---|---|---|---|---|---|
| **iter 1 *l*=1** | $top[1]$ | $\omega_1$ | $-2\pi(0.0)$ | | $-2\pi(0.0)$ |
| | $top[2]$ | $\omega_2$ | $-2\pi(0.00)$ | | $-2\pi(0.01)$ |
| | $top[3]$ | $\omega_3$ | $-2\pi(0.000)$ | | $-2\pi(0.011)$ |
| | ⋮ | ⋮ | ⋮ | | ⋮ |
| | $top[n_{top}-2]$ | $\omega_{n_{top}-2}$ | $-2\pi(0.\underbrace{00\cdots0}_{n_{top}-3})$ | | $-2\pi(0.\underbrace{01\cdots1}_{n_{top}-3})$ |
| | $top[n_{top}-1]$ | $\omega_{n_{top}-1}$ | $-2\pi(0.\underbrace{00\cdots0}_{n_{top}-2})$ | | $-2\pi(0.\underbrace{01\cdots1}_{n_{top}-2})$ |
| | $top[n_{top}]$ | $\omega_{n_{top}}$ | $-2\pi(0.\underbrace{00\cdots0}_{n_{top}-1})$ | | $-2\pi(0.\underbrace{01\cdots1}_{n_{top}-1})$ |

Generally, the goal of the *l*-th iteration is to extract the information of the low $l\times n_{top}$ bits of the eigenvalues $\phi_j$. Taking the case in Fig. 3(a) as an example, we have $n^{iter\ l-1}_{nonzero}=t$ when $d_{t-1}\leq(l-1)\times n_{top}\leq d_t$, which means that $t$ divergences in the eigenvalues can be differentiated after $l-1$ iterations. The high $q$ bits of $bin_q^{iter\ l}$ in the $l$-th iteration are designed in the same way as in Table 1. As mentioned above, the low $(l-1)\times n_{top}$ bits of $bin_q^{iter\ l}$ in the $l$-th iteration take the same values as those of $\phi_{j'}(P^{total}_{j'\phi_{j'}_iter\ (l-1)\times n_{top}} \neq 0)$ extracted after $l-1$ iterations in HIPEA-3, and they have $t$ possible values corresponding to $t$ divergences. Therefore, there are $t\times 2^{n_{top}-1}$ values to be traversed for $bin_{n_{top}}^{iter\ l}$ in the $l$-th iteration, and the number of experiments to be conducted is also $t\times 2^{n_{top}-1}$ in this iteration. In analogy with HIPEA-1, measuring probabilities $P^{total}_{j'\phi_{j'}_iter\ l\times n_{top}}$ are obtained after the $l$-th iteration in HIPEA-3, and the low $l\times n_{top}$ bits of the eigenvalues $\phi_j$ can be extracted accordingly.

According to Lemma 1, after $\lceil m/n_{top}\rceil$ ($\lceil\quad\rceil$ means rounding up to an integer) iterations, all the eigenvalues $\phi_j$ and the absolute values of the corresponding projection coefficients $|\beta_j|$ can be extracted.

(2) The absolute value $|u_{jp}|$ of each element of the normalized eigenvector is extracted similarly as in HIPEA-1.

(3) The values of $\beta_j u_j$ are extracted similarly as in HIPEA-1.

(4) The solution of the linear systems of equations is obtained as $x=\Sigma_j \beta_j u_j/\phi_j$.

In HIPEA-3, the number of ancillary qubits is flexible, and the number of iterations is reduced to $\lceil m/n_{top}\rceil$. As the cost of "flexible number of ancillary qubits", the number of experiments in each iteration increases in order to traverse all possible combinations of rotation parameters. The divergences may appear in any iteration in HIPEA-3, two extreme cases are considered here to analyze the total required number of experiments:

**Case 1** $n_{nonzero}^{iter\ 1}=n_{nonzero}^{iter\ 2}=\cdots=n_{nonzero}^{iter\ \lceil m/n_{top}\rceil}=N$ ($N$ divergences can be differentiated or $N$ non-zero measuring possibilities are obtained in the first iteration): The number of experiments required is the largest, reaching $2^{n_{top}-1}+N\times(\lceil m/n_{top}\rceil-1)\times 2^{n_{top}-1}$;

**Case 2** $n_{nonzero}^{iter\ 1}=n_{nonzero}^{iter\ 2}=\cdots=n_{nonzero}^{iter\ \lceil m/n_{top}\rceil-1}=1$, $n_{nonzero}^{iter\ \lceil m/n_{top}\rceil}=N$ ( $N$ divergences can be differentiated or $N$ non-zero measuring possibilities are obtained in the last iteration): The number of experiments required is the smallest, which is $\lceil m/n_{top}\rceil\times(2^{n_{top}-1}-1)$.

In terms of qubit resource consumption, the number of qubits in the top register can be greater or less than $N$ in HIPEA-3, which has greater flexibility than in HIPEA-2.

## 4. Results

In this section, three HIPEA algorithms are employed in solving linear equations $Ax=b$ with $A$ and $b$ being

$$A=2^{-9}\times\begin{bmatrix}224.5271 & -0.7218 & -0.0960 & -0.4480\\ -0.7218 & 127.8316 & -0.8596 & -0.1073\\ -0.0960 & -0.8596 & 125.9457 & -0.0794\\ -0.4480 & -0.1073 & -0.0794 & 189.6956\end{bmatrix} \tag{27}$$

$$b=[0.3538\ \ -0.5054\ \ 0.0396\ \ 0.7860]^T \tag{28}$$

The eigenvalues of the Hermitian matrix $A$ are shown in Eq. (20) and Fig. 3(b), and the corresponding eigenvectors $u_1$, $u_2$, $u_3$ and $u_4$ are

$$u_1=\begin{bmatrix}u_{11}\\u_{12}\\u_{13}\\u_{14}\end{bmatrix}=\begin{bmatrix}-0.7444\\0.1296\\-0.0496\\0.6531\end{bmatrix},\ u_2=\begin{bmatrix}u_{21}\\u_{22}\\u_{23}\\u_{24}\end{bmatrix}=\begin{bmatrix}0.3976\\0.6253\\0.5593\\0.3716\end{bmatrix},\ u_3=\begin{bmatrix}u_{31}\\u_{32}\\u_{33}\\u_{34}\end{bmatrix}=\begin{bmatrix}0.5356\\-0.3225\\-0.4458\\0.6406\end{bmatrix},\ u_4=\begin{bmatrix}u_{41}\\u_{42}\\u_{43}\\u_{44}\end{bmatrix}=\begin{bmatrix}-0.0295\\-0.6987\\0.6971\\0.1581\end{bmatrix} \quad (29)$$

Decomposing the normalized column vector $b$ on the eigenspace of $A$ spanned by these eigenvectors, we can obtain the corresponding projection coefficients as $\beta_1=0.1825$ ,$\beta_2=0.1389$, $\beta_3=0.8384$ and $\beta_4=0.4945$ .The exact solution $x$ obtained by the classic algorithm is taken as a benchmark in the analysis:

$$x=A^{-1}b=[4.9584\ \ -1.1007\ \ -0.6430\ \ 6.2099]^T \quad (30)$$

The three HIPEA algorithms are implemented in PyQPanda, an open-source quantum computing software development framework. For clarity, $\tilde{\beta}$, $\tilde{u}_j$ and $\tilde{x}$ are used instead in the following to indicate the results by the three HIPEA algorithms. $P_{\phi_{j(m-l+1)}\phi_{j(m-l+2)}\cdots\phi_{jm}}$ is defined as the probability that the measuring results of qubits in the top register is $\phi_{j(m-l+1)}\phi_{j(m-l+2)}\cdots\phi_{jm}$ after the *l*-th iteration, and eigenvalues different from Eq. (20) are marked as $\phi_0$ with measuring possibilities $P_{0\phi_0_iter\ l}^{total}=0$ .

### *4.1. Results of HIPEA-1*

According to the quantum circuits in Fig. 4, the specific settings of HIPEA-1 in this example are: (1) Two qubits are allocated to the bottom register to map the 4-dimensional column vector $b$; (2) Since each eigenvalue can be represented by one nine-bit binary number, nine iterations are needed to estimate these eigenvalues; (3) As shown in Fig. 3(b), there are three divergences in total in the estimation process, so four experiments are required accordingly to differentiate the eigenvalues.

The HIPEA-1 algorithm described in Sec. 3.1 is applied to the example, and results in each step are as follows:

(1) After nine iterations of the four experiments, the four eigenvalues in Eq. (20) are extracted from the measurement results of the top register. The probabilities of obtaining these four eigenvalues are respectively

$$\begin{aligned}P_{101110000}&=P_{1\phi_1_iter\ 9}^{total}=\tilde{\beta}_1^2=0.0344, \quad P_{000010110}=P_{2\phi_2_iter\ 9}^{total}=\tilde{\beta}_2^2=0.01975\\P_{011011011}&=P_{3\phi_3_iter\ 9}^{total}=\tilde{\beta}_3^2=0.24440, \quad P_{000111011}=P_{4\phi_4_iter\ 9}^{total}=\tilde{\beta}_4^2=0.70386\end{aligned} \quad (31)$$

leading to

$$|\tilde{\beta}_1|=0.1828,\ |\tilde{\beta}_2|=0.1405,\ |\tilde{\beta}_3|=0.4944,\ |\tilde{\beta}_4|=0.8390 \quad (32)$$

The rotation parameters in each iteration of the four experiments are shown in Table A.1 together with the measuring probabilities.

(2) The measurement is performed on the bottom register when the measurement results of the top register are, respectively, $\phi_1$, $\phi_2$, $\phi_3$ and $\phi_4$, and the results are

$$\begin{aligned}|\tilde{u}_1|&=[0.7522\ \ 0.1352\ \ 0.0349\ \ 0.6440]^T, \quad |\tilde{u}_2|=[0.4079\ \ 0.6213\ \ 0.5500\ \ 0.3809]^T\\|\tilde{u}_3|&=[0.0202\ \ 0.6991\ \ 0.6975\ \ 0.1551]^T, \quad |\tilde{u}_4|=[0.5365\ \ 0.3240\ \ 0.4435\ \ 0.6407]^T\end{aligned} \quad (33)$$

(3) According to Eq. (23), the following values of $\tilde{\beta}_j\tilde{u}_j$ are obtained:

$$\begin{aligned}\tilde{\beta}_1\tilde{u}_1&=[-0.1375\ \ 0.0247\ \ -0.0064\ \ 0.1177]^T, \quad \tilde{\beta}_2\tilde{u}_2=[0.0573\ \ 0.0873\ \ 0.0773\ \ 0.0535]^T\\\tilde{\beta}_3\tilde{u}_3&=[-0.0100\ \ -0.3456\ \ 0.3448\ \ 0.0767]^T, \quad \tilde{\beta}_4\tilde{u}_4=[0.4501\ \ -0.2718\ \ -0.3721\ \ 0.5375]^T\end{aligned} \quad (34)$$

(4) With the information in Eq. (20) and Eq. (34), the solution can be derived

$$\tilde{x} = \Sigma_j \tilde{\beta}_j \tilde{u}_j / \phi_j = [5.0255 \ \ -1.1005 \ \ -0.6328 \ \ 6.2534]^T \tag{35}$$

Comparing the obtained solution $\tilde{x}$ in Eq. (35) with the exact solution $x$ in Eq. (30), the relative error is

$$\epsilon = \frac{\|x - \tilde{x}\|}{\|x\|} = 0.0100 \tag{36}$$

With HIPEA-1 to solve the example linear systems of equations, a total of four experiments are conducted, and there are nine iterations in each experiment. In terms of qubit resource consumption, a total of three qubits are used, with one ancillary qubits in the top register and two qubits in the bottom register.

### *4.2. Results of HIPEA-2*

According to the quantum circuits in Fig. 5, the specific settings of HIPEA-2 in this example are: (1) Two qubits are allocated to the bottom register; (2) Nine iterations are used to estimate the eigenvalues; (3) As shown in Fig. 3(b), the three divergences in the process of eigenvalue estimation appear in the first, second and sixth iterations. Accordingly, two, three and four qubits in the top register are, respectively, put into use from the second, third and seventh iterations.

The HIPEA-2 described in Sec. 3.2 is applied to the example and results in each step are as follows: (1) After nine iterations, the four eigenvalues in Eq. (20) are extracted from the measurement results of the top register. The probabilities of obtaining these four eigenvalues are, respectively:

$$\begin{aligned} &P_{101110000} = P^{total}_{1\phi_1_iter\ 9} = \tilde{\beta}_1^2 = 0.03352, \quad P_{000010110} = P^{total}_{2\phi_2_iter\ 9} = \tilde{\beta}_2^2 = 0.02007 \\ &P_{011011011} = P^{total}_{3\phi_3_iter\ 9} = \tilde{\beta}_3^2 = 0.24276, \quad P_{000111011} = P^{total}_{4\phi_4_iter\ 9} = \tilde{\beta}_4^2 = 0.70180 \end{aligned} \tag{37}$$

leading to

$$|\tilde{\beta}_1| = 0.1831, \ |\tilde{\beta}_2| = 0.1417, \ |\tilde{\beta}_3| = 0.4927, \ |\tilde{\beta}_4| = 0.8377 \tag{38}$$

The rotation parameters and measuring probabilities in each iteration are shown in Table A.2.
(2) The measurement is performed on the bottom register when the measurement results of the top register are, respectively, $\phi_1$, $\phi_2$, $\phi_3$ and $\phi_4$, and the results are

$$\begin{aligned} &|\tilde{u}_1| = [0.7427 \ \ 0.1372 \ \ 0.0424 \ \ 0.6541]^T, \ |\tilde{u}_2| = [0.3973 \ \ 0.6268 \ \ 0.5579 \ \ 0.3717]^T \\ &|\tilde{u}_3| = [0.0349 \ \ 0.6974 \ \ 0.6979 \ \ 0.1595]^T, \ |\tilde{u}_4| = [0.5358 \ \ 0.3165 \ \ 0.4460 \ \ 0.6434]^T \end{aligned} \tag{39}$$

(3) According to Eq. (23), the following values of $\tilde{\beta}_j \tilde{u}_j$ are obtained:

$$\begin{aligned} &\tilde{\beta}_1 \tilde{u}_1 = [-0.1360 \ \ 0.0251 \ \ -0.0078 \ \ 0.1198]^T, \ \tilde{\beta}_2 \tilde{u}_2 = [0.0563 \ \ 0.0888 \ \ 0.0790 \ \ 0.0527]^T \\ &\tilde{\beta}_3 \tilde{u}_3 = [-0.0172 \ \ -0.3436 \ \ 0.3439 \ \ 0.0786]^T, \ \tilde{\beta}_4 \tilde{u}_4 = [0.4489 \ \ -0.2651 \ \ -0.3736 \ \ 0.5380]^T \end{aligned} \tag{40}$$

(4) With the information in Eq. (20) and Eq. (40), the solution is derived as

$$\tilde{x} = \Sigma_j \tilde{\beta}_j \tilde{u}_j / \phi_j = [4.9757 \ \ -1.0027 \ \ -0.6098 \ \ 6.2533]^T \tag{41}$$

Comparing the obtained solution $\tilde{x}$ in Eq. (41) with the exact solution $x$ in Eq. (30), the relative error is

$$\epsilon = \frac{\|x-\tilde{x}\|}{\|x\|} = 0.0141 \tag{42}$$

With HIPEA-2 to solve this example linear systems of equations, only one experiment with nine iterations is conducted. In terms of qubit resource consumption, a total of six qubits are used, with four ancillary qubits in the top register and two qubits in the bottom register.

### 4.3. Results of HIPEA-3

According to the quantum circuits in Fig. 6, the specific settings of HIPEA-3 in this example are: (1) Two qubits are allocated to the bottom register; (2) The number of qubits in the top register in HIPEA-3 is flexible, which is set to be $n_{top} = 3$ here for demonstration; (3) Since the eigenvalues can be represented by $m = 9$ digit binary numbers, $\lceil m/n_{top} \rceil = 3$ iterations are needed for eigenvalue estimation.

The HIPEA-3 described in Sec. 3.3 is applied to the example and results in each step are as follows: (1) After three iterations, the four eigenvalues in Eq. (20) are extracted from the measurement results of the top register. The probabilities of obtaining these four eigenvalues are, respectively:

$$\begin{aligned} P_{101110000} = P^{total}_{1\phi_1_iter\ 9} = \tilde{\beta}_1^2 = 0.03361,\ P_{000010110} = P^{total}_{2\phi_2_iter\ 9} = \tilde{\beta}_2^2 = 0.01892 \\ P_{011011011} = P^{total}_{3\phi_3_iter\ 9} = \tilde{\beta}_3^2 = 0.24549,\ P_{000111011} = P^{total}_{4\phi_4_iter\ 9} = \tilde{\beta}_4^2 = 0.70108 \end{aligned} \tag{43}$$

leading to

$$\left|\tilde{\beta}_1\right| = 0.1833,\ \left|\tilde{\beta}_2\right| = 0.1375,\ \left|\tilde{\beta}_3\right| = 0.4955,\ \left|\tilde{\beta}_4\right| = 0.8373 \tag{44}$$

The rotation parameters and measuring probabilities in each iteration are shown in Table A.3.
(2) The measurement is performed on the bottom register when the measurement results of the top register are, respectively, $\phi_1$, $\phi_2$, $\phi_3$ and $\phi_4$, and the results are

$$\begin{aligned} |\tilde{u}_1| = [0.7339\ \ 0.1295\ \ 0.0629\ \ 0.6639]^T,\ |\tilde{u}_2| = [0.4018\ \ 0.6263\ \ 0.5617\ \ 0.3618]^T \\ |\tilde{u}_3| = [0.0287\ \ 0.6984\ \ 0.6977\ \ 0.1568]^T,\ |\tilde{u}_4| = [0.5355\ \ 0.3227\ \ 0.4433\ \ 0.6423]^T \end{aligned} \tag{45}$$

(3) According to Eq. (23), the following values of $\tilde{\beta}_j \tilde{u}_j$ are obtained:

$$\begin{aligned} \tilde{\beta}_1\tilde{u}_1 = [-0.1345\ \ 0.0237\ \ -0.0115\ \ 0.1217]^T,\ \tilde{\beta}_2\tilde{u}_2 = [0.0553\ \ \ 0.0861\ \ \ 0.0773\ \ \ 0.0498]^T \\ \tilde{\beta}_3\tilde{u}_3 = [-0.0142\ \ -0.3460\ \ 0.3457\ \ 0.0777]^T,\ \tilde{\beta}_3\tilde{u}_3 = [0.4484\ \ -0.2702\ \ -0.3712\ \ 0.5378]^T \end{aligned} \tag{46}$$

(4) With the information in Eq. (20) and Eq. (46), the solution is derived as

$$\tilde{x} = \Sigma_j \tilde{\beta}_j \tilde{u}_j / \phi_j = [4.9568\ \ -1.1158\ \ -0.6308\ \ 6.1782]^T \tag{47}$$

Comparing the obtained solution $\tilde{x}$ in Eq. (47) with the exact solution $x$ in Eq. (30), the relative error is

$$\epsilon = \frac{\|x-\tilde{x}\|}{\|x\|} = 0.0048 \tag{48}$$

With HIPEA-3 to solve this example linear systems of equations, only three iterations are required, and a total of 32 experiments are conducted. In terms of qubit resource consumption, a total of five qubits are used, with three ancillary qubits in the top register and two qubits in the bottom register.

## 5. Conclusions

In this paper, three hybrid classical-quantum algorithms are designed based on the iterative phase estimation algorithm to solve linear systems of equations $Ax = b$, and less qubit resources are required in these hybrid algorithms than in quantum HHL algorithm. The hybrid algorithms store the measurement results of each iteration of each experiment in the classic register, and the non-zero measuring probabilities of states of the ancillary qubits in the top register are used to extract the eigenvalues of $A$ and the corresponding projection coefficients $\beta_j$ in $b = \Sigma_j \beta_j u_j$. When measuring with post-selection, the absolute value $|u_{jp}|$ of each element of the normalized eigenvector $u_j$ of $A$ can be obtained. The final solution will then be derived after the values of $\beta_j u_j$ in $b = \Sigma_j \beta_j u_j$ are determined.

The number of qubits in the bottom register is the same in our three hybrid algorithms, which is determined by the dimension of the column vector $b$, while the number of ancillary qubits in the top register is different. HIPEA-1 requires the least number of ancillary qubits (i.e. 1), and the eigenvalues of matrix are differentiated by conducting different experiments with $N$ iterations. HIPEA-2 achieves the same effect with multiple ancillary qubits (i.e. $N$) instead of multiple experiments, and the complexity just moves from “the number of experiments” to “the number of ancillary qubits”. HIPEA-3 differentiates different eigenvalues with flexible multiple ancillary qubits using a traversing strategy, and a set of experiments is performed in each iteration to traverse one bit by one bit all possible combinations of rotation parameters and thus all possible binary number strings of eigenvalues. The three algorithms are implemented in PyQPanda, and one specific example is employed to prove the feasibility of these algorithms when the eigenvalues of $A$ can be perfectly represented by finite binary number strings. The developed HIPEA algorithms with limited qubit resources broadens the application range of quantum computation in solving linear systems of equations.

In the future studies, it is worthy to modify HIPEA algorithms or develop new hybrid algorithms to solve $Ax = b$ when the eigenvalues of $A$ can not be perfectly represented by finite binary number strings. Moreover, how to calibrate the signs of $\beta_j u_{jp}$ with a more universal and rigorous scheme is also a question worth considering. In case that the eigenvalues and eigenvectors of the coefficient matrix $A$ are known in advance, we do not need to design redundant rotation parameters and only have to focus on how to extract the coefficients $\beta_j$ efficiently. Since modules of initial state preparation, Hamiltonian simulation and accurate measurement are included in HIPEA, the improvement schemes of these modules may possibly help to improve the performance of HIPEA.

## Appendix A. Rotation parameters and measuring probabilities of three HIPEA algorithms for one specific example

Table A.1 Rotation parameters and measuring probabilities of HIPEA-1

| Experiment 1 | Rotation parameters | Measuring Probabilities | |
|---|---|---|---|
| Iteration 1 | $\omega_1 = -2\pi(0.0)$ | $P_0 = P_{1\phi_1_iter1}^{total} = 0.05257$ | $P_1 = P_{3\phi_3_iter1}^{total} = 0.94742$ |
| Iteration 2 | $\omega_2 = -2\pi(0.00)$ | $P_{00} = P_{1\phi_1_iter2}^{total} = 0.03238$ | $P_{10} = P_{2\phi_2_iter2}^{total} = 0.01898$ |
| Iteration 3 | $\omega_3 = -2\pi(0.000)$ | $P_{000} = P_{1\phi_1_iter3}^{total} = 0.1012$ | $P_{100} = P_{0\phi_0_iter3}^{total} = 0$ |
| Iteration 4 | $\omega_4 = -2\pi(0.0000)$ | $P_{0000} = P_{1\phi_1_iter4}^{total} = 0.03387$ | $P_{1000} = P_{0\phi_0_iter4}^{total} = 0$ |
| Iteration 5 | $\omega_5 = -2\pi(0.00000)$ | $P_{00000} = P_{0\phi_0_iter5}^{total} = 0$ | $P_{10000} = P_{1\phi_1_iter5}^{total} = 0.03372$ |
| Iteration 6 | $\omega_6 = -2\pi(0.010000)$ | $P_{010000} = P_{0\phi_0_iter6}^{total} = 0$ | $P_{110000} = P_{1\phi_1_iter6}^{total} = 0.03234$ |
| Iteration 7 | $\omega_7 = -2\pi(0.0110000)$ | $P_{0110000} = P_{0\phi_0_iter7}^{total} = 0$ | $P_{1110000} = P_{1\phi_1_iter7}^{total} = 0.03345$ |
| Iteration 8 | $\omega_8 = -2\pi(0.01110000)$ | $P_{01110000} = P_{1\phi_1_iter8}^{total} = 0.03245$ | $P_{11110000} = P_{0\phi_0_iter8}^{total} = 0$ |
| Iteration 9 | $\omega_9 = -2\pi(0.001110000)$ | $P_{001110000} = P_{0\phi_0_iter9}^{total} = 0$ | $P_{101110000} = P_{1\phi_1_iter9}^{total} = 0.0344$ |
| **Experiment 2** | **Rotation parameters** | **Measuring Probabilities** | |
| Iteration 1 | $\omega_1 = -2\pi(0.0)$ | $P_0 = P_{1\phi_1_iter1}^{total} = 0.05257$ | $P_1 = P_{3\phi_3_iter1}^{total} = 0.94742$ |
| Iteration 2 | $\omega_2 = -2\pi(0.00)$ | $P_{00} = P_{1\phi_1_iter2}^{total} = 0.03238$ | $P_{10} = P_{2\phi_2_iter2}^{total} = 0.01898$ |
| Iteration 3 | $\omega_3 = -2\pi(0.010)$ | $P_{010} = P_{0\phi_0_iter3}^{total} = 0$ | $P_{110} = P_{2\phi_2_iter3}^{total} = 0.01943$ |
| Iteration 4 | $\omega_4 = -2\pi(0.0110)$ | $P_{0110} = P_{2\phi_2_iter4}^{total} = 0.01857$ | $P_{1110} = P_{0\phi_0_iter4}^{total} = 0$ |
| Iteration 5 | $\omega_5 = -2\pi(0.00110)$ | $P_{00110} = P_{0\phi_0_iter5}^{total} = 0$ | $P_{10110} = P_{2\phi_2_iter5}^{total} = 0.01917$ |
| Iteration 6 | $\omega_6 = -2\pi(0.010110)$ | $P_{010110} = P_{2\phi_2_iter6}^{total} = 0.01919$ | $P_{110110} = P_{0\phi_0_iter6}^{total} = 0$ |
| Iteration 7 | $\omega_7 = -2\pi(0.0010110)$ | $P_{0010110} = P_{2\phi_2_iter7}^{total} = 0.01969$ | $P_{1010110} = P_{0\phi_0_iter7}^{total} = 0$ |
| Iteration 8 | $\omega_8 = -2\pi(0.00010110)$ | $P_{00010110} = P_{2\phi_2_iter8}^{total} = 0.001933$ | $P_{10010110} = P_{0\phi_0_iter8}^{total} = 0$ |
| Iteration 9 | $\omega_9 = -2\pi(0.000010110)$ | $P_{000010110} = P_{2\phi_2_iter9}^{total} = 0.01975$ | $P_{100010110} = P_{0\phi_0_iter9}^{total} = 0$ |
| **Experiment 3** | **Rotation parameters** | **Measuring Probabilities** | |
| Iteration 1 | $\omega_1 = -2\pi(0.0)$ | $P_0 = P_{1\phi_1_iter1}^{total} = 0.05257$ | $P_1 = P_{3\phi_3_iter1}^{total} = 0.94742$ |
| Iteration 2 | $\omega_2 = -2\pi(0.01)$ | $P_{01} = P_{0\phi_0_iter2}^{total} = 0$ | $P_{11} = P_{3\phi_3_iter2}^{total} = 0.94763$ |
| Iteration 3 | $\omega_3 = -2\pi(0.011)$ | $P_{011} = P_{3\phi_3_iter3}^{total} = 0.94703$ | $P_{111} = P_{0\phi_0_iter3}^{total} = 0$ |
| Iteration 4 | $\omega_4 = -2\pi(0.0011)$ | $P_{0011} = P_{0\phi_0_iter4}^{total} = 0$ | $P_{1011} = P_{3\phi_3_iter4}^{total} = 0.94680$ |
| Iteration 5 | $\omega_5 = -2\pi(0.01011)$ | $P_{01011} = P_{0\phi_0_iter5}^{total} = 0$ | $P_{11011} = P_{3\phi_3_iter5}^{total} = 0.94644$ |
| Iteration 6 | $\omega_6 = -2\pi(0.011011)$ | $P_{011011} = P_{3\phi_3_iter6}^{total} = 0.24140$ | $P_{111011} = P_{4\phi_4_iter6}^{total} = 0.70658$ |
| Iteration 7 | $\omega_7 = -2\pi(0.0011011)$ | $P_{0011011} = P_{0\phi_0_iter7}^{total} = 0$ | $P_{1011011} = P_{3\phi_3_iter7}^{total} = 0.24466$ |
| Iteration 8 | $\omega_8 = -2\pi(0.01011011)$ | $P_{01011011} = P_{0\phi_0_iter8}^{total} = 0$ | $P_{11011011} = P_{3\phi_3_iter8}^{total} = 0.24543$ |
| Iteration 9 | $\omega_9 = -2\pi(0.011011011)$ | $P_{011011011} = P_{3\phi_3_iter9}^{total} = 0.24440$ | $P_{111011011} = P_{0\phi_0_iter9}^{total} = 0$ |
| **Experiment 4** | **Rotation parameters** | **Measuring Probabilities** | |
| Iteration 1 | $\omega_1 = -2\pi(0.0)$ | $P_0 = P_{1\phi_1_iter1}^{total} = 0.05257$ | $P_1 = P_{3\phi_3_iter1}^{total} = 0.94742$ |
| Iteration 2 | $\omega_2 = -2\pi(0.01)$ | $P_{01} = P_{0\phi_0_iter2}^{total} = 0$ | $P_{11} = P_{3\phi_3_iter2}^{total} = 0.94763$ |
| Iteration 3 | $\omega_3 = -2\pi(0.011)$ | $P_{011} = P_{3\phi_3_iter3}^{total} = 0.94703$ | $P_{111} = P_{0\phi_0_iter3}^{total} = 0$ |
| Iteration 4 | $\omega_4 = -2\pi(0.0011)$ | $P_{0011} = P_{0\phi_0_iter4}^{total} = 0$ | $P_{1011} = P_{3\phi_3_iter4}^{total} = 0.94680$ |
| Iteration 5 | $\omega_5 = -2\pi(0.01011)$ | $P_{01011} = P_{0\phi_0_iter5}^{total} = 0$ | $P_{11011} = P_{3\phi_3_iter5}^{total} = 0.94644$ |
| Iteration 6 | $\omega_6 = -2\pi(0.011011)$ | $P_{011011} = P_{3\phi_3_iter6}^{total} = 0.24140$ | $P_{111011} = P_{4\phi_4_iter6}^{total} = 0.70658$ |
| Iteration 7 | $\omega_7 = -2\pi(0.0111011)$ | $P_{0111011} = P_{4\phi_4_iter7}^{total} = 0.70275$ | $P_{1111011} = P_{0\phi_0_iter7}^{total} = 0$ |
| Iteration 8 | $\omega_8 = -2\pi(0.00111011)$ | $P_{00111011} = P_{4\phi_4_iter8}^{total} = 0.70406$ | $P_{10111011} = P_{0\phi_0_iter8}^{total} = 0$ |
| Iteration 9 | $\omega_9 = -2\pi(0.000111011)$ | $P_{000111011} = P_{4\phi_4_iter9}^{total} = 0.70386$ | $P_{100111011} = P_{0\phi_0_iter9}^{total} = 0$ |

Table A.2 Rotation parameters and measuring probabilities of HIPEA-2.
Blank indicates that the corresponding qubit in the top register is not in use in this iteration.

| top[1] | Rotation parameters | Measuring Probabilities | |
|---|---|---|---|
| Iteration 1 | $\omega_1^1 = -2\pi(0.0)$ | $P_0 = P^{total}_{1\phi_1_iter\,1} = 0.05258$ | $P_1 = P^{total}_{3\phi_3_iter\,1} = 0.94742$ |
| Iteration 2 | $\omega_2^1 = -2\pi(0.00)$ | $P_{00} = P^{total}_{1\phi_1_iter\,2} = 0.03379$ | $P_{10} = P^{total}_{2\phi_2_iter\,2} = 0.01902$ |
| Iteration 3 | $\omega_3^1 = -2\pi(0.000)$ | $P_{000} = P^{total}_{1\phi_1_iter\,3} = 0.0335$ | $P_{100} = P^{total}_{0\phi_0_iter\,3} = 0$ |
| Iteration 4 | $\omega_4^1 == -2\pi(0.0000)$ | $P_{0000} = P^{total}_{1\phi_1_iter\,4} = 0.03270$ | $P_{1000} = P^{total}_{0\phi_0_iter\,4} = 0$ |
| Iteration 5 | $\omega_5^1 == -2\pi(0.00000)$ | $P_{00000} = P^{total}_{0\phi_0_iter\,5} = 0$ | $P_{10000} = P^{total}_{1\phi_1_iter\,5} = 0.03371$ |
| Iteration 6 | $\omega_6^1 == -2\pi(0.010000)$ | $P_{010000} = P^{total}_{0\phi_0_iter\,6} = 0$ | $P_{110000} = P^{total}_{1\phi_1_iter\,6} = 0.03274$ |
| Iteration 7 | $\omega_7^1 == -2\pi(0.0110000)$ | $P_{0110000} = P^{total}_{0\phi_0_iter\,7} = 0$ | $P_{1110000} = P^{total}_{1\phi_1_iter\,7} = 0.03345$ |
| Iteration 8 | $\omega_8^1 == -2\pi(0.01110000)$ | $P_{01110000} = P^{total}_{1\phi_1_iter\,8} = 0.03306$ | $P_{11110000} = P^{total}_{0\phi_0_iter\,8} = 0$ |
| Iteration 9 | $\omega_9^1 == -2\pi(0.001110000)$ | $P_{001110000} = P^{total}_{0\phi_0_iter\,9} = 0$ | $P_{101110000} = P^{total}_{1\phi_1_iter\,9} = 0.03352$ |
| top[2] | Rotation parameters | Measuring Probabilities | |
| Iteration 1 | | | |
| Iteration 2 | $\omega_2^2 == -2\pi(0.01)$ | $P_{01} = P^{total}_{0\phi_0_iter\,2} = 0$ | $P_{11} = P^{total}_{3\phi_3_iter\,2} = 0.94764$ |
| Iteration 3 | $\omega_3^2 = -2\pi(0.011)$ | $P_{011} = P^{total}_{3\phi_3_iter\,3} = 0.94797$ | $P_{111} = P^{total}_{0\phi_0_iter\,3} = 0$ |
| Iteration 4 | $\omega_4^2 = -2\pi(0.0011)$ | $P_{0011} = P^{total}_{0\phi_0_iter\,4} = 0$ | $P_{1011} = P^{total}_{3\phi_3_iter\,4} = 0.94754$ |
| Iteration 5 | $\omega_5^2 = -2\pi(0.01011)$ | $P_{01011} = P^{total}_{0\phi_0_iter\,5} = 0$ | $P_{11011} = P^{total}_{3\phi_3_iter\,5} = 0.94758$ |
| Iteration 6 | $\omega_6^2 = -2\pi(0.011011)$ | $P_{011011} = P^{total}_{3\phi_3_iter\,6} = 0.24589$ | $P_{111011} = P^{total}_{4\phi_4_iter\,6} = 0.7024$ |
| Iteration 7 | $\omega_7^2 = -2\pi(0.0011011)$ | $P_{0011011} = P^{total}_{0\phi_0_iter\,7} = 0$ | $P_{1011011} = P^{total}_{3\phi_3_iter\,7} = 0.24422$ |
| Iteration 8 | $\omega_8^2 = -2\pi(0.01011011)$ | $P_{01011011} = P^{total}_{0\phi_0_iter\,8} = 0$ | $P_{11011011} = P^{total}_{3\phi_3_iter\,8} = 0.2452$ |
| Iteration 9 | $\omega_9^2 = -2\pi(0.011011011)$ | $P_{011011011} = P^{total}_{3\phi_3_iter\,9} = 0.24276$ | $P_{111011011} = P^{total}_{0\phi_0_iter\,9} = 0$ |
| top[3] | Rotation parameters | Measuring Probabilities | |
| Iteration 1 | | | |
| Iteration 2 | | | |
| Iteration 3 | $\omega_3^3 = -2\pi(0.010)$ | $P_{010} = P^{total}_{0\phi_0_iter\,3} = 0$ | $P_{110} = P^{total}_{2\phi_2_iter\,3} = 0.01907$ |
| Iteration 4 | $\omega_4^3 = -2\pi(0.0110)$ | $P_{0110} = P^{total}_{2\phi_2_iter\,4} = 0.01906$ | $P_{1110} = P^{total}_{0\phi_0_iter\,4} = 0$ |
| Iteration 5 | $\omega_5^3 = -2\pi(0.00110)$ | $P_{00110} = P^{total}_{0\phi_0_iter\,5} = 0$ | $P_{10110} = P^{total}_{2\phi_2_iter\,5} = 0.01954$ |
| Iteration 6 | $\omega_6^3 = -2\pi(0.010110)$ | $P_{010110} = P^{total}_{2\phi_2_iter\,6} = 0.01957$ | $P_{110110} = P^{total}_{0\phi_0_iter\,6} = 0$ |
| Iteration 7 | $\omega_7^3 = -2\pi(0.0010110)$ | $P_{0010110} = P^{total}_{2\phi_2_iter\,7} = 0.01965$ | $P_{1010110} = P^{total}_{0\phi_0_iter\,7} = 0$ |
| Iteration 8 | $\omega_8^3 = -2\pi(0.00010110)$ | $P_{00010110} = P^{total}_{2\phi_2_iter\,8} = 0.70406$ | $P_{10010110} = P^{total}_{0\phi_0_iter\,8} = 0$ |
| Iteration 9 | $\omega_9^3 = -2\pi(0.000010110)$ | $P_{000010110} = P^{total}_{2\phi_2_iter\,9} = 0.02007$ | $P_{100010110} = P^{total}_{0\phi_0_iter\,9} = 0$ |
| top[4] | Rotation parameters | Measuring Probabilities | |
| Iteration 1 | | | |
| Iteration 2 | | | |
| Iteration 3 | | | |
| Iteration 4 | | | |
| Iteration 5 | | | |
| Iteration 6 | | | |
| Iteration 7 | $\omega_7^4 = -2\pi(0.0111011)$ | $P_{0111011} = P^{total}_{4\phi_4_iter\,7} = 0.70208$ | $P_{1111011} = P^{total}_{0\phi_0_iter\,7} = 0$ |
| Iteration 8 | $\omega_8^4 = -2\pi(0.00111011)$ | $P_{00111011} = P^{total}_{4\phi_4_iter\,8} = 0.7039$ | $P_{10111011} = P^{total}_{0\phi_0_iter\,8} = 0$ |
| Iteration 9 | $\omega_9^4 = -2\pi(0.000111011)$ | $P_{000111011} = P^{total}_{4\phi_4_iter\,9} = 0.7018$ | $P_{100111011} = P^{total}_{0\phi_0_iter\,9} = 0$ |

Table A.3 Rotation parameters and measuring probabilities of HIPEA-3

(a) The first iteration

| | Experiment 1 | | Experiment 2 | |
|---|---|---|---|---|
| Iteration 1 | $\omega_1 = -2\pi(0.0)$ | | $\omega_1 = -2\pi(0.0)$ | |
| | $\omega_2 = -2\pi(0.00)$ | | $\omega_2 = -2\pi(0.01)$ | |
| | $\omega_3 = -2\pi(0.000)$ | | $\omega_3 = -2\pi(0.001)$ | |
| Measuring Probabilities | $P_{000} = P^{total}_{1\phi_1_iter3} = 0.03362$ | $P_{100} = P^{total}_{0\phi_0_iter3} = 0$ | $P_{001} = P^{total}_{0\phi_0_iter3} = 0$ | $P_{101} = P^{total}_{0\phi_0_iter3} = 0$ |
| | Experiment 3 | | Experiment 4 | |
| Iteration 1 | $\omega_1 = -2\pi(0.0)$ | | $\omega_1 = -2\pi(0.0)$ | |
| | $\omega_2 = -2\pi(0.00)$ | | $\omega_2 = -2\pi(0.00)$ | |
| | $\omega_3 = -2\pi(0.010)$ | | $\omega_3 = -2\pi(0.011)$ | |
| Measuring Probabilities | $P_{010} = P^{total}_{0\phi_0_iter3} = 0$ | $P_{110} = P^{total}_{2\phi_2_iter3} = 0.01931$ | $P_{011} = P^{total}_{3\phi_3_iter3} = 0.94844$ | $P_{111} = P^{total}_{0\phi_0_iter3} = 0$ |

(b) The second iteration

| | Experiment 5 | | Experiment 6 | |
|---|---|---|---|---|
| Iteration 2 | $\omega_4 = -2\pi(0.0000)$ | | $\omega_4 = -2\pi(0.0000)$ | |
| | $\omega_5 = -2\pi(0.00000)$ | | $\omega_5 = -2\pi(0.01000)$ | |
| | $\omega_6 = -2\pi(0.000000)$ | | $\omega_6 = -2\pi(0.001000)$ | |
| Measuring Probabilities | $P_{000000} = P^{total}_{0\phi_0_iter6} = 0$ | $P_{100000} = P^{total}_{0\phi_0_iter6} = 0$ | $P_{001000} = P^{total}_{0\phi_0_iter6} = 0$ | $P_{101000} = P^{total}_{0\phi_0_iter6} = 0$ |
| | Experiment 7 | | Experiment 8 | |
| Iteration 2 | $\omega_4 = -2\pi(0.0000)$ | | $\omega_4 = -2\pi(0.0000)$ | |
| | $\omega_5 = -2\pi(0.00000)$ | | $\omega_5 = -2\pi(0.01000)$ | |
| | $\omega_6 = -2\pi(0.010000)$ | | $\omega_6 = -2\pi(0.011000)$ | |
| Measuring Probabilities | $P_{010000} = P^{total}_{0\phi_0_iter6} = 0$ | $P_{110000} = P^{total}_{1\phi_1_iter6} = 0.0325$ | $P_{011000} = P^{total}_{0\phi_0_iter6} = 0$ | $P_{111000} = P^{total}_{0\phi_0_iter6} = 0$ |
| | Experiment 9 | | Experiment 10 | |
| Iteration 2 | $\omega_4 = -2\pi(0.0110)$ | | $\omega_4 = -2\pi(0.0110)$ | |
| | $\omega_5 = -2\pi(0.00110)$ | | $\omega_5 = -2\pi(0.01110)$ | |
| | $\omega_6 = -2\pi(0.000110)$ | | $\omega_6 = -2\pi(0.001110)$ | |
| Measuring Probabilities | $P_{000100} = P^{total}_{0\phi_0_iter6} = 0$ | $P_{100110} = P^{total}_{0\phi_0_iter6} = 0$ | $P_{001110} = P^{total}_{0\phi_0_iter6} = 0$ | $P_{101110} = P^{total}_{0\phi_0_iter6} = 0$ |
| | Experiment 11 | | Experiment 12 | |
| Iteration 2 | $\omega_4 = -2\pi(0.0110)$ | | $\omega_4 = -2\pi(0.0110)$ | |
| | $\omega_5 = -2\pi(0.00110)$ | | $\omega_5 = -2\pi(0.01110)$ | |
| | $\omega_6 = -2\pi(0.010110)$ | | $\omega_6 = -2\pi(0.011110)$ | |
| Measuring Probabilities | $P_{010110} = P^{total}_{2\phi_2_iter6} = 0.0201$ | $P_{110110} = P^{total}_{0\phi_0_iter6} = 0$ | $P_{456789} = P^{total}_{0\phi_0_iter6} = 0$ | $P_{110110} = P^{total}_{0\phi_0_iter6} = 0$ |
| | Experiment 13 | | Experiment 14 | |
| Iteration 2 | $\omega_4 = -2\pi(0.0011)$ | | $\omega_4 = -2\pi(0.0011)$ | |
| | $\omega_5 = -2\pi(0.00011)$ | | $\omega_5 = -2\pi(0.01011)$ | |
| | $\omega_6 = -2\pi(0.000011)$ | | $\omega_6 = -2\pi(0.001011)$ | |
| Measuring Probabilities | $P_{010110} = P^{total}_{0\phi_0_iter6} = 0$ | $P_{110110} = P^{total}_{0\phi_0_iter6} = 0$ | $P_{010110} = P^{total}_{0\phi_0_iter6} = 0$ | $P_{110110} = P^{total}_{0\phi_0_iter6} = 0$ |
| | Experiment 15 | | Experiment 16 | |
| Iteration 2 | $\omega_4 = -2\pi(0.0011)$ | | $\omega_4 = -2\pi(0.0011)$ | |
| | $\omega_5 = -2\pi(0.00011)$ | | $\omega_5 = -2\pi(0.01011)$ | |
| | $\omega_6 = -2\pi(0.010011)$ | | $\omega_6 = -2\pi(0.011011)$ | |
| Measuring Probabilities | $P_{011110} = P^{total}_{0\phi_0_iter6} = 0$ | $P_{111110} = P^{total}_{0\phi_0_iter6} = 0$ | $P_{011011} = P^{total}_{3\phi_3_iter6} = 0.24537$ | $P_{111011} = P^{total}_{4\phi_4_iter6} = 0.70272$ |

(c) The third iteration

| | | | | |
|---|---|---|---|---|
| | **Experiment 17** | | **Experiment 18** | |
| **Iteration 3** | $\omega_7 = -2\pi(0.0110000)$<br>$\omega_8 = -2\pi(0.00110000)$<br>$\omega_9 = -2\pi(0.000110000)$ | | $\omega_7 = -2\pi(0.0110000)$<br>$\omega_8 = -2\pi(0.01110000)$<br>$\omega_9 = -2\pi(0.001110000)$ | |
| **Measuring Probabilities** | $P_{000110000} = P^{total}_{0\phi_0_iter9} = 0$ | $P_{100110000} = P^{total}_{0\phi_0_iter9} = 0$ | $P_{001110000} = P^{total}_{0\phi_0_iter9} = 0$ | $P_{101110000} = P^{total}_{1\phi_1_iter9} = 0.03361$ |
| | **Experiment 19** | | **Experiment 20** | |
| **Iteration 3** | $\omega_7 = -2\pi(0.0110000)$<br>$\omega_8 = -2\pi(0.00110000)$<br>$\omega_9 = -2\pi(0.010110000)$ | | $\omega_7 = -2\pi(0.0110000)$<br>$\omega_8 = -2\pi(0.01110000)$<br>$\omega_9 = -2\pi(0.011110000)$ | |
| **Measuring Probabilities** | $P_{010110000} = P^{total}_{0\phi_0_iter9} = 0$ | $P_{110110000} = P^{total}_{0\phi_0_iter9} = 0$ | $P_{011110000} = P^{total}_{0\phi_0_iter9} = 0$ | $P_{111110000} = P^{total}_{0\phi_0_iter9} = 0$ |
| | **Experiment 21** | | **Experiment 22** | |
| **Iteration 3** | $\omega_7 = -2\pi(0.0010110)$<br>$\omega_8 = -2\pi(0.00010110)$<br>$\omega_9 = -2\pi(0.000010110)$ | | $\omega_7 = -2\pi(0.0010110)$<br>$\omega_8 = -2\pi(0.01010110)$<br>$\omega_9 = -2\pi(0.001010110)$ | |
| **Measuring Probabilities** | $P_{000010110} = P^{total}_{2\phi_2_iter9} = 0.01892$ | $P_{100010110} = P^{total}_{0\phi_0_iter9} = 0$ | $P_{001010110} = P^{total}_{0\phi_0_iter9} = 0$ | $P_{101010110} = P^{total}_{0\phi_0_iter9} = 0$ |
| | **Experiment 23** | | **Experiment 24** | |
| **Iteration 3** | $\omega_7 = -2\pi(0.0010110)$<br>$\omega_8 = -2\pi(0.00010110)$<br>$\omega_9 = -2\pi(0.010010110)$ | | $\omega_7 = -2\pi(0.0010110)$<br>$\omega_8 = -2\pi(0.01010110)$<br>$\omega_9 = -2\pi(0.011010110)$ | |
| **Measuring Probabilities** | $P_{001010110} = P^{total}_{0\phi_0_iter9} = 0$ | $P_{101010110} = P^{total}_{0\phi_0_iter9} = 0$ | $P_{011010110} = P^{total}_{0\phi_0_iter9} = 0$ | $P_{111010110} = P^{total}_{0\phi_0_iter9} = 0$ |
| | **Experiment 25** | | **Experiment 26** | |
| **Iteration 3** | $\omega_7 = -2\pi(0.0011011)$<br>$\omega_8 = -2\pi(0.00011011)$<br>$\omega_9 = -2\pi(0.000011011)$ | | $\omega_7 = -2\pi(0.0011011)$<br>$\omega_8 = -2\pi(0.01011011)$<br>$\omega_9 = -2\pi(0.001011011)$ | |
| **Measuring Probabilities** | $P_{000011011} = P^{total}_{0\phi_0_iter9} = 0$ | $P_{100011011} = P^{total}_{0\phi_0_iter9} = 0$ | $P_{001011011} = P^{total}_{0\phi_0_iter9} = 0$ | $P_{101011011} = P^{total}_{0\phi_0_iter9} = 0$ |
| | **Experiment 27** | | **Experiment 28** | |
| **Iteration 3** | $\omega_7 = -2\pi(0.0011011)$<br>$\omega_8 = -2\pi(0.00011011)$<br>$\omega_9 = -2\pi(0.010011011)$ | | $\omega_7 = -2\pi(0.0011011)$<br>$\omega_8 = -2\pi(0.01011011)$<br>$\omega_9 = -2\pi(0.011011011)$ | |
| **Measuring Probabilities** | $P_{010011011} = P^{total}_{0\phi_0_iter9} = 0$ | $P_{110011011} = P^{total}_{0\phi_0_iter9} = 0$ | $P_{011011011} = P^{total}_{3\phi_3_iter9} = 0.24549$ | $P_{111011011} = P^{total}_{0\phi_0_iter9} = 0$ |
| | **Experiment 29** | | **Experiment 31** | |
| **Iteration 3** | $\omega_7 = -2\pi(0.0111011)$<br>$\omega_8 = -2\pi(0.00111011)$<br>$\omega_9 = -2\pi(0.000111011)$ | | $\omega_7 = -2\pi(0.0111011)$<br>$\omega_8 = -2\pi(0.01111011)$<br>$\omega_9 = -2\pi(0.001111011)$ | |
| **Measuring Probabilities** | $P_{000111011} = P^{total}_{4\phi_4_iter9} = 0.70108$ | $P_{100111011} = P^{total}_{0\phi_0_iter9} = 0$ | $P_{001111011} = P^{total}_{0\phi_0_iter9} = 0$ | $P_{101111011} = P^{total}_{0\phi_0_iter9} = 0$ |
| | **Experiment 31** | | **Experiment 32** | |
| **Iteration 3** | $\omega_7 = -2\pi(0.0111011)$<br>$\omega_8 = -2\pi(0.00111011)$<br>$\omega_9 = -2\pi(0.010111011)$ | | $\omega_7 = -2\pi(0.0111011)$<br>$\omega_8 = -2\pi(0.01111011)$<br>$\omega_9 = -2\pi(0.011111011)$ | |
| **Measuring Probabilities** | $P_{010111011} = P^{total}_{0\phi_0_iter9} = 0$ | $P_{110111011} = P^{total}_{0\phi_0_iter9} = 0$ | $P_{011111011} = P^{total}_{0\phi_0_iter9} = 0$ | $P_{111111011} = P^{total}_{0\phi_0_iter9} = 0$ |

## References

[1] X.D. Cai, C. Weedbrook, Z.E. Su, M.C. Chen, M. Gu, M.J. Zhu, L. Li, N. Le Liu, C.Y. Lu, J.W. Pan, Experimental quantum computing to solve systems of linear equations, Phys. Rev. Lett. 110 (2013) 1–5. https://doi.org/10.1103/PhysRevLett.110.230501.

[2] A.W. Harrow, A. Hassidim, S. Lloyd, Quantum algorithm for linear systems of equations, Phys. Rev. Lett. 103 (2009). https://doi.org/10.1103/PhysRevLett.103.150502.

[3] P. Schleich, How to solve a linear system of equations using a quantum computer, Semin. Proj. (2019) 1–35. www.mathcces.rwth-aachen.de/_media/3teaching/00projects/schleich.pdf.

[4] C. Shao, Reconsider HHL algorithm and its related quantum machine learning algorithms, ArXiv. (2018) 5.

[5] J. Dickens, Quantum Computing Algorithms for Applied Linear Algebra Author, (2019).

[6] A. Carrera Vázquez, S. Wörner, R. Hiptmair, Quantum Algorithm for Solving Tri-Diagonal Linear Systems of Equations, (2018) 1–24.

[7] B. Duan, J. Yuan, C.H. Yu, J. Huang, C.Y. Hsieh, A survey on HHL algorithm: From theory to application in quantum machine learning, Phys. Lett. Sect. A Gen. At. Solid State Phys. 384 (2020) 126595. https://doi.org/10.1016/j.physleta.2020.126595.

[8] J. Preskill, Quantum computing in the NISQ era and beyond, Quantum. 2 (2018) 1–20. https://doi.org/10.22331/q-2018-08-06-79.

[9] Y. Lee, J. Joo, S. Lee, Hybrid quantum linear equation algorithm and its experimental test on IBM Quantum Experience, Sci. Rep. 9 (2019) 1–12. https://doi.org/10.1038/s41598-019-41324-9.

[10] V. Bužek, R. Derka, S. Massar, Optimal quantum clocks, Asymptot. Theory Quantum Stat. Inference Sel. Pap. (2005) 477–486. https://doi.org/10.1142/9789812563071_0032.

[11] K.M. Svore, M.B. Hastings, M. Freedman, Faster phase estimation, Quantum Inf. Comput. 14 (2014) 306–328.

[12] R. Cleve, A. Ekert, C. Macchiavello, M. Mosca, Quantum algorithms revisited, Proc. R. Soc. A Math. Phys. Eng. Sci. 454 (1998) 339–354. https://doi.org/10.1098/rspa.1998.0164.

[13] X.Q. Zhou, P. Kalasuwan, T.C. Ralph, J.L. O'brien, Calculating unknown eigenvalues with a quantum algorithm, Nat. Photonics. 7 (2013) 223–228. https://doi.org/10.1038/nphoton.2012.360.

[14] V. Parasa, M. Perkowski, Quantum phase estimation using multivalued logic, Proc. - 41st IEEE Int. Symp. Mult. Logic, ISMVL 2011. (2011) 224–229. https://doi.org/10.1109/ISMVL.2011.47.

[15] T.E. O'Brien, B. Tarasinski, B.M. Terhal, Quantum phase estimation of multiple eigenvalues for small-scale (noisy) experiments, ArXiv. (2018).

[16] T.E. O'Brien, B. Tarasinski, B.M. Terhal, Quantum phase estimation of multiple eigenvalues for small-scale (noisy) experiments, New J. Phys. 21 (2019). https://doi.org/10.1088/1367-2630/aafb8e.

[17] C.J. O'Loan, Iterative phase estimation, J. Phys. A Math. Theor. 43 (2010).

[18] M. Dobšíček, G. Johansson, V. Shumeiko, G. Wendin, Arbitrary accuracy iterative quantum phase estimation algorithm using a single ancillary qubit: A two-qubit benchmark, Phys. Rev. A - At. Mol. Opt. Phys. 76 (2007) 1–4. https://doi.org/10.1103/PhysRevA.76.030306.

[19] X.M. Liu, J. Luo, X.P. Sun, Experimental realization of arbitrary accuracy iterative phase estimation algorithms on ensemble quantum computers, Chinese Phys. Lett. 24 (2007) 3316–3319. https://doi.org/10.1088/0256-307X/24/12/007.

[20] V. Giovannetti, S. Lloyd, L. MacCone, Quantum random access memory, Phys. Rev. Lett. 100 (2008) 1–4. https://doi.org/10.1103/PhysRevLett.100.160501.

[21] V. Giovannetti, S. Lloyd, L. MacCone, Architectures for a quantum random access memory, Phys. Rev. A - At. Mol. Opt. Phys. 78 (2008) 1–9. https://doi.org/10.1103/PhysRevA.78.052310.

[22] A.M. Childs, N. Wiebe, Hamiltonian simulation using linear combinations of unitary operations, Quantum Inf. Comput. 12 (2012) 901–924. https://doi.org/10.26421/qic12.11-12-1.

[23] D.W. Berry, A.M. Childs, Black-box hamiltonian simulation and unitary implementation, Quantum Inf. Comput. 12 (2012) 29–62. https://doi.org/10.26421/QIC12.1-2.

[24] M.A. Nielsen, M.J. Bremner, J.L. Dodd, A.M. Childs, C.M. Dawson, Universal simulation of Hamiltonian dynamics for quantum systems with finite-dimensional state spaces, Phys. Rev. A - At. Mol. Opt. Phys. 66 (2002) 1–12. https://doi.org/10.1103/PhysRevA.66.022317.

[25] G.H. Low, I.L. Chuang, Optimal Hamiltonian Simulation by Quantum Signal Processing, Phys. Rev. Lett. 118 (2017) 1–5. https://doi.org/10.1103/PhysRevLett.118.010501.

[26] R. Santagati, J. Wang, A.A. Gentile, S. Paesani, N. Wiebe, J.R. McClean, S. Morley-Short, P.J. Shadbolt, D. Bonneau, J.W. Silverstone, D.P. Tew, X. Zhou, J.L. O'Brien, M.G. Thompson, Witnessing eigenstates for quantum simulation of Hamiltonian spectra, Sci. Adv. 4 (2018) 1–12. https://doi.org/10.1126/sciadv.aap9646.

[27] D.W. Berry, G. Ahokas, R. Cleve, B.C. Sanders, Efficient quantum algorithms for simulating sparse hamiltonians, Commun. Math. Phys. 270 (2007) 359–371. https://doi.org/10.1007/s00220-006-0150-x.

[28] S. Aaronson, Quantum computing, postselection, and probabilistic polynomial-time, Proc. R. Soc. A Math. Phys. Eng. Sci. 461 (2005) 3473–3482. https://doi.org/10.1098/rspa.2005.1546.

[29] B.D. Clader, B.C. Jacobs, C.R. Sprouse, Preconditioned quantum linear system algorithm, Phys. Rev. Lett. 110 (2013) 1–5. https://doi.org/10.1103/PhysRevLett.110.250504.

[30] D. Dervovic, M. Herbster, P. Mountney, S. Severini, N. Usher, L. Wossnig, Quantum linear systems algorithms: A primer, ArXiv. (2018).